\documentclass[aps,pra,twocolumn,a4paper,showpacs,nofootinbib]{revtex4}
\usepackage{amsmath}
\usepackage{amssymb}
\usepackage{amsfonts}
\usepackage{times,txfonts}
\usepackage{color}
\usepackage{bm}
\usepackage{graphicx}
\usepackage{epsfig}
\usepackage{epstopdf}
\usepackage{verbatim}

\newcommand{\bra}[1]{\langle #1|}
\newcommand{\ket}[1]{|#1\rangle}
\newcommand{\braket}[2]{\langle #1|#2\rangle}
\newcommand{\expec}[1]{\langle #1 \rangle}

\begin{document}
\title{Power-law behavior in the quantum-resonant evolution of the $\delta$-kicked accelerator}

\author{P. L. Halkyard}
\affiliation{Department of Physics, Durham University, Rochester Building, South Road, Durham DH1 3LE, United Kingdom}

\author{M. Saunders}
\affiliation{Department of Physics, Durham University, Rochester Building, South Road, Durham DH1 3LE, United Kingdom}

\author{K. J. Challis}
\affiliation{Lundbeck Foundation Theoretical Center for Quantum System Research, Department of Physics and Astronomy, University of Aarhus, DK-8000 {\AA}rhus C, Denmark}

\author{S. A. Gardiner}
\affiliation{Department of Physics, Durham University, Rochester Building, South Road, Durham DH1 3LE, United Kingdom}

\begin{abstract}
We consider the atom-optical $\delta$-kicked accelerator when the initial momentum distribution is symmetric.  We demonstrate the existence of quantum-resonant dynamics, and derive analytic expressions for the system evolution.  In particular, we consider the dynamical evolution of the momentum moments and find that all even-ordered momentum moments exhibit a power law growth.  In the ultracold (zero-temperature) limit the exponent is determined by the order of the moment, whereas for a broad, thermal initial momentum distribution the exponent is reduced by one.  To demonstrate the power law behavior explicitly we consider the evolutions of the second- and fourth-order momentum moments, and cumulants, for an initially Gaussian momentum distribution corresponding to the Maxwell-Boltzmann distribution of an ideal gas at thermal equilibrium. 
\end{abstract}
\date{\today}
\maketitle

\section{Introduction}
Quantum resonances and antiresonances \cite{Casati1979,Izrailev1979,Izrailev1980,Dana1994,Dana1995,Dana1996,Dana2005,Dana2006a,Dana2006b,Wimberger2003,Wimberger2004,Wimberger2005a,Lepers2008,Bharucha1999,Oskay2000,Sadgrove2004,Duffy2004b,Kanem2007,Currivan2008,Deng1999} are striking signatures of ordered quantum manifestations in the $\delta$-kicked rotor, a paradigm system in the study of classical \cite{Lichtenberg1992} and quantum chaotic dynamics \cite{Reichl2004,Haake2001,Gutzwiller1990}.  The atom-optical realization of the $\delta$-kicked rotor (more accurately denoted the $\delta$-kicked particle) has in recent years proved to be a fertile testing ground for quantum-chaotic phenomena \cite{Oskay2000,Sadgrove2004,Bharucha1999,Moore1995,Ammann1998a,dArcy2001b,Moore1994,Klappauf1999,Steck2000,Milner2000,Oskay2003,Vant2000,Doherty2000,Kanem2007,Currivan2008,dArcy2003,Duffy2004b,Ryu2006,Szriftgiser2002,Ammann1998b,Vant1999,Williams2004,Duffy2004a,Tonyushkin2008, Wu2008}.  An important variant is the $\delta$-kicked accelerator, where the kicking potential is aligned parallel to the local gravitational acceleration \cite{Oberthaler1999,Godun2000,dArcy2001a,Schlunk2003a,Schlunk2003b,Ma2004,Buchleitner2006,Behinaein2006,Fishman2002,Fishman2003,Bach2005,Guarneri2008}.  This gives rise to phenomena closely related to quantum resonances, for example quantum accelerator modes, and fractional quantum resonances \cite{Saunders2008}.  Such atom-optical realizations involve subjecting a cold, dilute atomic gas to a periodically-pulsed laser standing wave.  The amplitude, phase, periodicity, and duration of the pulses can be controlled to a high degree of precision, allowing theoretical predictions to be thoroughly investigated. 

In this paper we investigate the quantum-resonant behavior \cite{Dana1995,Dana1996,Dana2005,Dana2006a,Dana2006b,Wimberger2003,Wimberger2004,Wimberger2005a,Lepers2008,Bharucha1999,Oskay2000,Sadgrove2004,Duffy2004b,Kanem2007,Currivan2008} of laser-driven clouds of freely-falling, laser-cooled atoms.  Such a system closely models the $\delta$-kicked accelerator, of which the $\delta$-kicked particle is a particular case.  Experimentally, the system dynamics are typically interrogated by measuring the atomic centre-of mass momentum distribution, using a time-of-flight technique \cite{dArcy2001a}.  Although there are practical resolution limits, it is then in principle possible to determine all momentum moments from the measured distribution.  A limited number of moments is frequently sufficient to characterize the system: the second-order momentum moment provides a clear signature to distinguish between quantum-resonant dynamics and dynamical localization in the $\delta$-kicked particle \cite{Saunders2007}, whereas the fourth-order moment is necessary to distinguish between different fractional quantum resonances in the $\delta$-kicked accelerator at finite temperature \cite{Saunders2008}.  The momentum moments therefore provide conceptually simple measures to characterize the effect of quantum resonant dynamics on the atomic centre-of-mass momentum distribution. 
The main focus of this paper is to provide fully analytical predictions of the momentum-moment dynamics of the atom-optical $\delta$-kicked accelerator under certain physically motivated conditions.  We present detailed derivations of the key analytical results utilized in Saunders \textit{et al}.\ \cite{Saunders2007,Saunders2008}.

Assuming only that the initial momentum distribution is symmetric, we find that the time evolution of all even momentum moments follow power-law behaviors, and examine how such behavior differs between opposing limits of the initial distribution being either extremely narrow or broad.  Assuming a Maxwell-Boltzmann distribution for an ideal gas, we determine explicit analytic expressions for the time evolution of the second- and fourth-order momentum moments when the gas is initially in an ultracold limit, or at finite temperature.  We also outline a procedure for calculating higher-order momentum moments for a given (but arbitrary) initial momentum distribution.  Although we neglect two-body interactions, we note that our model can also be used to describe Bose-Einstein condensates where the scattering length has been made negligibly small by tuning around a Feshbach resonance \cite{Inouye1998,Roberts1998,Kohler2006,Gustavsson2008}.  

The paper is organized as follows. 
In Sec.\ \ref{sec2} we sketch our theoretical model for the atom-optical $\delta$-kicked accelerator, derive the time evolution operator from which the time evolution of the momentum moments can be determined, and derive general conditions for which antiresonance and resonance occur.  
In Sec.\ \ref{sec3} we consider only momentum distributions that are initially symmetric, and discuss the power-law behavior of the evolution of the even-ordered momentum moments in the two limits of narrow and broad initial momentum distributions.  In Sec.\ \ref{sec4} we consider the time evolution of the second- and fourth-order momentum moments and cumulants for momentum distributions that are initially Gaussian, that is, corresponding to a finite temperature ideal gas. Section \ref{sec5} consists of the conclusions, which are then  followed by five technical appendices.

\section{Atom-optical $\delta$-kicked accelerator\label{sec2}} 
\subsection{System Hamiltonian\label{sec2b}}
We consider a cloud of trapped and laser-cooled alkali atoms.  The cloud is released from the trap and addressed by an appropriate configuration of off-resonant laser beams forming a vertically aligned, pulsed, sinusoidal potential.  We neglect atomic interactions and treat the cloud as an ensemble of single-particle systems with Hamiltonian \cite{dArcy2001a,Saunders2008}
\begin{equation}
\begin{split}
\hat{H} = \frac{\hat{p}^2}{2M} + Ma\hat{z}-\frac{\hbar\Omega_R^2}{8\Delta}\cos(K\hat{z})f(t). \label{hatom}
\end{split} 
\end{equation}
Here $\hat{p}$ and $\hat{z}$ are the centre-of-mass momentum and position operators, $\Omega_R$ is the Rabi frequency, $K/2$ is the laser wavenumber along the $z$ direction, $M$ is the mass, $\Delta$ is the detuning and $f(t)$ describes the periodic pulses.  The parameter $a$ is the relative acceleration between the atomic cloud and the optical standing-wave potential and is in general given by $a\equiv g-a_{\phi}$, where $g$ is the local gravitational acceleration and $a_\phi$ can be chosen with high precision by tuning the relative phase of the laser beams creating the optical field \cite{Saunders2008}.  When $a$ is set to zero, we recover the atom-optical $\delta$-kicked rotor \cite{Saunders2007}.

We assume the pulse duration to be sufficiently short for the atoms to be in the Raman-Nath regime, i.e., the displacement of the atoms during the pulse is significantly less than the wavelength of the standing wave.  Hence, to a good approximation, the pulses may be modeled as a train of $\delta$ functions \cite{Saunders2007} giving the quantum $\delta$-kicked accelerator Hamiltonian \cite{dArcy2001a}
\begin{equation}
\begin{split}
\hat{H}_{\delta\text{ka}} = \frac{\hat{p}^2}{2M} + Ma\hat{z}-\hbar\phi_d\cos(K\hat{z})\sum_{n=0}^{\infty}\delta(t-nT),\label{hdelta}
\end{split} 
\end{equation}
where $\phi_d \equiv \Omega_R^2t_p/8\Delta$, $t_p$ is the pulse duration, and $T$ is the pulse periodicity.  For convenience we apply the unitary operator $\hat{U} = \exp(iMa\hat{z}/\hbar)$ \cite{Bach2005,Fishman2002,Fishman2003}, to transform Hamiltonian (\ref{hdelta}) to the spatially periodic form:
\begin{equation}
\begin{split}
\tilde{H}_{\delta\text{ka}}  = \frac{(\hat{p}-Mat)^2}{2M} -\hbar\phi_d\cos(K\hat{z})\sum_{n=0}^{\infty}\delta(t-nT),
\end{split}
\label{hspat} 
\end{equation}
for which we can invoke Bloch theory \cite{Saunders2007,Kittel1996}.

\subsection{Time evolution}
\subsubsection{Transformed Floquet operator \label{sec2c}}
The $\delta$-kicked accelerator Hamiltonian (\ref{hdelta}) is periodic in time, and the system evolution can be described in terms of a Floquet (kick-to-kick time-evolution) operator.  The transformed Floquet operator corresponding to the transformed Hamiltonian (\ref{hspat}) is $\exp(-iMa^2T^3[3n^2-3n+1]/6\hbar) \tilde{F}_{n}$ \cite{Bach2005,Saunders2008}, where
\begin{equation}
\begin{split}
\tilde{F}_n = e^{-i\left[\hbar K^2(\hat{k}+\hat{\beta})^2T/2M - K a(\hat{k}+\hat{\beta})(2n-1)T^2/2\right]}e^{i\phi_d \cos(\hat{\theta})},
\end{split} 
\label{Fprime0}
\end{equation}
i.e., we absorb a global phase into the definition of $\tilde{F}_n$.\footnote{The global phase is corrected slightly from that given previously \cite{dArcy2001a}.}  We have separated the momentum and position operators into discrete and continuous components \cite{Bach2005}:
\begin{equation}
\hat{z} = K^{-1}(2\pi \hat{l}+\hat{\theta}), \quad \hat{p}  =  \hbar K(\hat{k}+\hat{\beta}),
\end{equation}
 where the eigenvalues of $\hat{l}$ and $\hat{k}$ are integers, and the eigenvalues of $\hat{\theta}$ and $\hat{\beta}$ are $\theta \in [-\pi,\pi)$ and $\beta \in [-1/2,1/2)$.  

Because the Hamiltonian (\ref{hspat}) is periodic in space, the quasimomentum $\beta$ is conserved \cite{Fishman2002,Fishman2003}, i.e., $\hat{\beta}$ commutes with $\tilde{H}_{\delta\text{ka}}$ \cite{Bach2005} and only momentum eigenstates with eigenvalues differing by integer multiples of $\hbar K$ are coupled \cite{dArcy2001a}.  Therefore, within a particular quasimomentum subspace, the dynamics are determined by
\begin{equation}
\begin{split}
\tilde{F}_n(\beta) = e^{-i\left[\hbar K^2(\hat{k}+\beta)^2T/2M - K a(\hat{k}+\beta)(2n-1)T^2/2\right]}e^{i\phi_d \cos(\hat{\theta})}.
\end{split} 
\label{Fprime}
\end{equation}

As is the case for the $\delta$-kicked rotor or particle, quantum resonances or antiresonances occur when $T=\ell T_T/2$ \cite{Saunders2007}, where $T_T = 2\pi M/\hbar K^2$ is the Talbot time \cite{Godun2000} (so-named in analogy with the Talbot effect in optics \cite{Hecht2002}).  In this case, Eq.\ (\ref{Fprime}) becomes
\begin{equation}
\begin{split}
\tilde{F}_n(\beta) = e^{-i\pi\beta[\beta\ell-\Omega(2n-1)]}e^{-i\pi[(\ell-\Omega)+2\ell\beta]\hat{k}}e^{i\phi_d \cos(\hat{\theta})},
\end{split} 
\label{F}
\end{equation}
where we have used $\exp(-i\ell\pi \hat{k}^2)=\exp(-i\ell\pi \hat{k})$, and introduced the dimensionless effective gravitational acceleration $\Omega=KaT^2/2\pi$ \cite{Buchleitner2006}.

\subsubsection{Time evolution of the momentum moments \label{sec2d}}
The time evolution of a general momentum eigenstate $\ket{k+\beta}$ can be derived by consecutively applying the transformed Floquet operator (\ref{F}), the details of which are given in Appendix \ref{matrix_element} [See Eqs.\ (\ref{Eq:p_amplitude}) and (\ref{eta_def})].  We deduce that
\begin{equation}
\begin{split}
\ket{\Psi(t=nT)} =& \tilde{F}_n(\beta)\tilde{F}_{n-1}(\beta)\ldots\tilde{F}_1(\beta)\ket{k+\beta}\\=& \sum_{j=-\infty}^{\infty}
J_{j-k} (\omega) 
e^{i(j-k)\chi}
e^{-i\pi n \alpha k}
e^{-in^{2}\pi(k+\beta)\Omega}
\\ &\times
e^{-in\pi\beta^{2}\ell}
\ket{j+\beta},
\end{split} 
\label{evolvedstate}
\end{equation}
where $\omega$ and $\chi$ are defined in Eqs.\ (\ref{phidnu}) and (\ref{mu_eqn}), and \footnote{The parameter $\alpha$ is related to $\Upsilon$ \cite{Saunders2007,Saunders2008} by $\Upsilon = \alpha \pi/2$.} 
\begin{equation}
\alpha = (1+2\beta)\ell.
\label{Eq:DefAlpha}
\end{equation}

Defining the dimensionless momentum $\hat{P} \equiv \hat{k} + \hat{\beta} \equiv \hat{p}/\hbar K$, it follows straightforwardly from Eq.\ (\ref{evolvedstate}) that the $m$-th order momentum moment of the evolved state is
\begin{equation}
\expec{\hat{P}^{m}}_n = \sum_{j=-\infty}^{\infty}J^2_{j-k}\left( \phi_d\eta\right)(j+\beta)^m,
\label{pm0}
\end{equation}
where $\eta\equiv |\omega|/\phi_{d}=|\tilde{\nu}|$, with
\begin{equation}
\tilde{\nu}  = \sum_{j=0}^{n-1}e^{i\pi (\alpha j - \Omega j^2)}. \label{eta}
\end{equation}
We note that Eq.\ (\ref{eta}) has the form of a Gauss sum, and that Gauss sums have applications in number theory and various areas of theoretical physics \cite{Bigourd2008,Gilowski2008,Apostol1976,Armitage2000}.  

By extension, the evolution of the moments for an initial statistical mixture of momentum eigenstates, with distribution $D_k(\beta)$, is described by \cite{Saunders2007,Saunders2008}
\begin{equation}
\expec{\hat{P}^m}_n=\int_{-1/2}^{1/2}d\beta \sum_{j,k=-\infty}^{\infty}J_{j-k}^2\left( \phi_d
\eta\right)(j+\beta)^m D_k(\beta). 
\label{pm}
\end{equation}

\subsection{Existence of resonances and antiresonances} \label{sec2e}
\subsubsection{Manifestation of resonances in $\eta$} \label{sec2e2}
Quantum resonances are characterized by unbounded growth in the system energy, and we see that this is due to the constructive inteference of oscillatory terms in Eq.\ (\ref{eta}).  They occur in the $\delta$-kicked rotor or particle when $\phi_{d}\eta$ in Eq.\ (\ref{evolvedstate}) can be replaced by $\phi_{d}n$ \cite{Saunders2007}. To achieve such resonant growth in  $\eta$, the oscillatory terms contained within the sum in Eq.\ (\ref{eta}) must add in phase.  It follows that the summand of $\tilde{\nu}$ must be periodic with respect to the counting index $j$.  The period $Q$ (if present) is determined from
\begin{equation}
e^{i\pi [\alpha (j+Q) - \Omega( j^{2} + 2jQ + Q^{2}) ] } = e^{i\pi (\alpha j - \Omega j^{2})},
\label{Eq:PrePeriodCondition}
\end{equation}
implying that
\begin{equation}
\alpha Q - 2j\Omega Q - \Omega Q^2 = 2A, 
\label{periodcondition0}
\end{equation}
must be fulfilled, where $A$ is an arbitrary integer. 

The period $Q$ should be independent of the counting index $j$, and so $\Omega Q$ must be integer.  Hence, $\Omega$ is rational, and if we set $\Omega= r/s$ ($r$ and $s$ are integers with no common factors), it follows that the smallest possible value for $Q$ is $s$.
With this information we deduce a simpler condition from Eq.\ (\ref{Eq:PrePeriodCondition}):
\begin{equation}
s(\alpha - r) = 2A. \label{rescondition}
\end{equation}
Eliminating $\alpha$ using Eq.\ (\ref{Eq:DefAlpha}) gives
\begin{equation}
\beta = \frac{2A+ s(r-\ell)}{2\ell s}, \label{reslocation}
\end{equation} 
indicating the quasimomentum values, subject to $\beta \in (-1/2,1/2]$, for which quantum resonances occur.  

\begin{figure}[tbp]
\centering
\includegraphics[width=8.8cm]{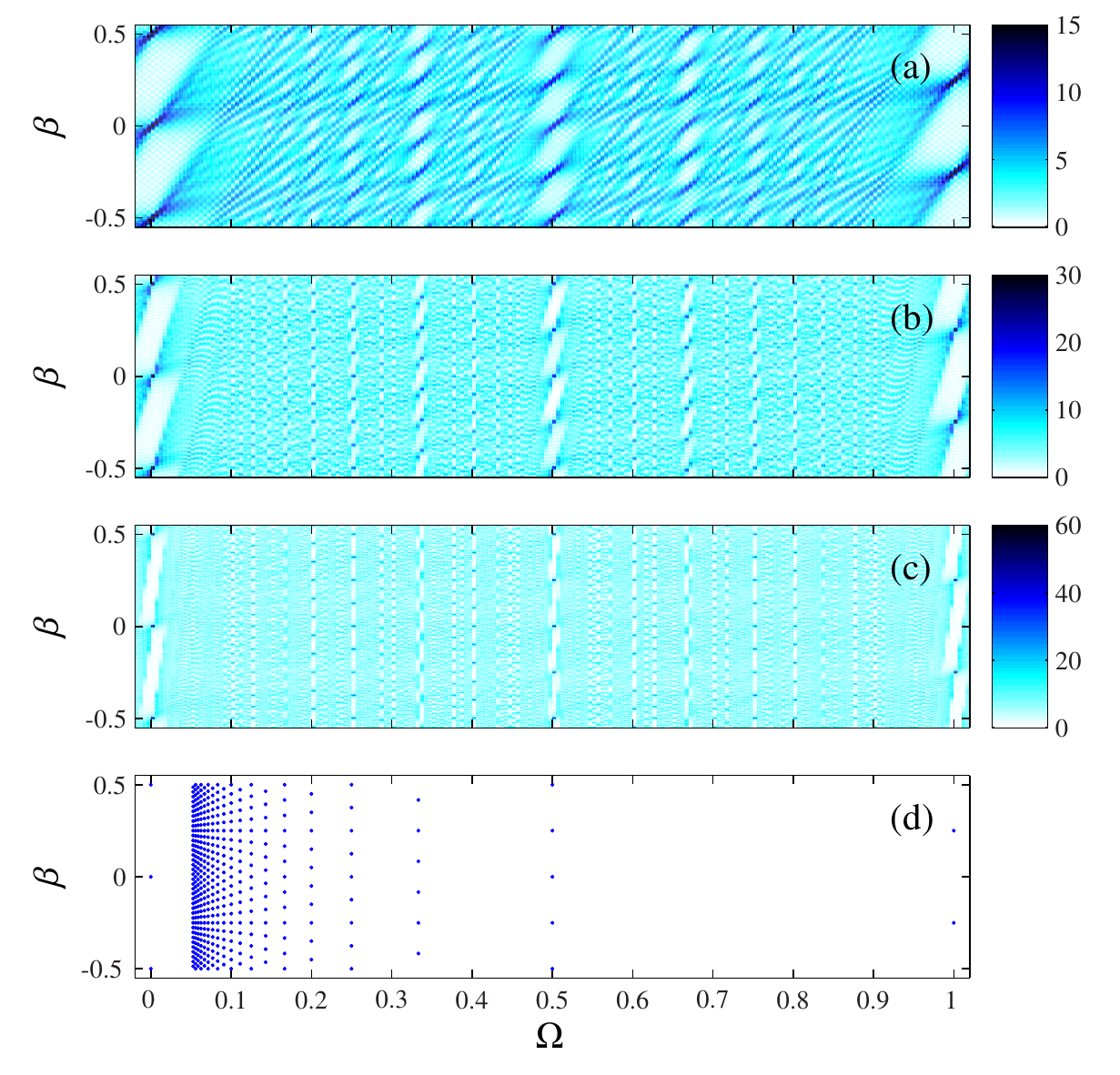}
\caption{(Color online)  
Parameter $\eta$ as a function of $\Omega$ and $\beta$ for $\ell=2$ at (a) $15$ kicks (b) $30$ kicks, and (c) $60$ kicks.  An antiresonance of order $2N_s$, with $\Omega=r/s$, will initially mimic exactly resonant growth for $N_ss$ kicks.  The diminishing finite size of the resonances in (a) -- (c) is due to nearby high-order antiresonances which become apparent as $n$ increases.  Panel (d) shows the location of those resonances (see Eq.\ (\ref{reslocation})) for $\Omega=1/s$, where $s=1,2,  \ldots , 20$.  The resonances become infinite in number as $\Omega \to 0$, but for $\Omega=0$ there are only three resonances.
}
\label{reslocationfig}
\end{figure}

\subsubsection{Manifestation of antiresonances in $\eta$}
An antiresonance, in the strictest sense, is when the state periodically cycles back to its initial condition \cite{Saunders2007}.  This can only occur when $\phi_{d}\eta$ is periodic in $n$, implying that $\exp(i\pi [\alpha j - \Omega j^{2}])$ should be oscillatory, i.e.,  
\begin{equation}
e^{i\pi [\alpha (j+Q) - \Omega( j^{2} + 2jQ + Q^{2}) ] } = -e^{i\pi (\alpha j - \Omega j^{2})}= e^{i\pi (\alpha j - \Omega j^{2}+1)}.
\label{antirescondition0}
\end{equation}
Proceeding along similar lines as for the resonant case (see Sec.\ \ref{sec2e2}) this implies that $Q=s$, and that
\begin{equation}
s(\alpha-r) = 2A+1, \label{antirescondition}
\end{equation}
where $A$ is an arbitrary integer.  Condition (\ref{antirescondition}) ensures that  $\eta = 0$ when $n$ is an even multiple of $s$.  

For a given value of $\Omega=r/s$, higher-order antiresonances also exist for different values of $\beta$ \cite{Saunders2007}.  These are described by the condition
\begin{equation}
\begin{split}
e^{i\pi [\alpha (j+Q) - (r/s)( j^{2} + 2jQ + Q^{2}) ] } = & e^{i\pi N_{r}/N_{s}}e^{i\pi [\alpha j - (r/s) j^{2}]}, \\
\Rightarrow
e^{i\pi [\alpha (j+N_{s}Q) - (r/s)( j^{2} + 2jN_{s}Q + N_{s}^{2}Q^{2}) ] } = & -e^{i\pi [\alpha j - (r/s) j^{2}]}.
\end{split}
\label{Nantirescondition0}
\end{equation}
As above, it follows that $Q=s$; hence, the revival period of the initial state is $2N_{s}s$. Condition (\ref{Nantirescondition0}) implies 
a generalization of Eq.\ (\ref{antirescondition}), i.e.,
\begin{equation}
s(\alpha-r) =  2A + \frac{N_{r}}{N_{s}}.
\label{Nantirescondition}
\end{equation}
Antiresonances with revival periods $2N_{s} s$ therefore occur for quasimomentum values (subject to $\beta \in (-1/2,1/2]$)
\begin{equation}
\beta = \frac{2A+ N_{r}/N_{s} + s(r-\ell)}{2\ell s}.
\label{antireslocation}
\end{equation}

How both resonances and antiresonances are manifest in $\eta$, and their dependence on $\Omega$ and $\beta$, is illustrated in Fig.\ \ref{reslocationfig}.  This demonstrates that the resonances are dense in the parameter space, but diminish in strength for increasing $s$.  Furthermore, resonances become more sharply defined for increasing $n$ as nearby high-order antiresonances become more important.

\section{Time evolution of the momentum moments \label{sec3}}
\subsection{General symmetric momentum distributions} \label{sec3b}
\subsubsection{Physical motivation}
In the atom-optical kicked systems under consideration, the atomic gas is initially held in a harmonic trapping potential.  The ground state of the system is therefore spatially symmetric, both in the presence and absence of significant interatomic interactions.  We consider the momentum moment dynamics for the case of symmetric initial momentum distributions $D(P)$, such that $D(P)=D(-P)$.  In particular, we provide a detailed analytic investigation of the second- and fourth-order momentum moments.  These moments are most relevant as signatures of dynamics in the atom optical delta-kicked particle \cite{Saunders2007} and accelerator \cite{Saunders2008}. The second-order momentum moment  [$m=2$ in Eq.\ (\ref{pm})] is proportional to the mean kinetic energy, and the fourth-order moment contains information about the degree to which the distribution is peaked.

\subsubsection{Consequences of symmetry \label{odd}}

We consider the initial state to be an incoherent ensemble of momentum eigenstates, i.e., the momentum representation of the density operator is diagonal. For a symmetric initial momentum distribution, the initial population of state $\ket{-k-\beta}$ is always equal to the initial population of state $\ket{k+\beta}$. 
Using Eq.\ (\ref{pm0}), and explicitly stating the $\beta$-dependence of $\eta(\beta)$, for an initial state $\ket{-k-\beta}$ the momentum moments evolve as
\begin{equation}
\expec{\hat{P}^{m}}_n 
= (-1)^{m}\sum_{j=-\infty}^{\infty}J^2_{j-k}( \phi_d\eta(-\beta))(j+\beta)^{m},
\label{p2m1}
\end{equation}
where we have relabelled the summation (setting $j$ to $-j$) and used that $J_{-j}(x) = (-1)^jJ_j(x)$ \cite{Abramowitz1964}.

Comparing Eqs.\ (\ref{p2m1}) and (\ref{pm0}), we see that when $\eta(-\beta)=\eta(\beta)$, it follows that
$\expec{\hat{P}^{2m+1}}_n$ for an initial state $\ket{-k-\beta}$ is equal to $-\expec{\hat{P}^{2m+1}}_n$ 
for an initial state $\ket{k+\beta}$. The momentum moment for a statistical mixture is the normalized sum of the moments of the individual component states.  Consequently, for initially symmetric momentum distributions, the odd moments are identically zero whenever $\eta(\beta)$ is \textit{in general\/} equal to $\eta(-\beta)$, i.e., for all values of the quasimomentum $\beta$.

\subsubsection{Invariance of $\eta$\label{Sec:Invariance}}

The invariance of $\eta(\beta)$ upon changing $\beta$ to $-\beta$ is a useful property which is often satisfied.  In particular, for rational $\Omega=r/s$ it is valid at resonance [see Eq.\ (\ref{rescondition})] when $n$ is a multiple of $s$, and at antiresonance when $s$ is an even multiple of $s$.  Here we derive these results.

To begin we write Eq.\ (\ref{eta}), changing the summation index $j$ to $n-j$, i.e.,
\begin{equation}
\begin{split}
\tilde{\nu}_{n+1}(\beta)
=& \sum_{j=0}^{n}e^{i\pi[(1+2\beta)\ell (n-j) -(r/s)(n^{2} - 2nj +j^{2})]}\\
=& e^{i\pi(1+2\beta)\ell n -(r/s)n^{2}}
\sum_{j=0}^{n}
e^{i\pi[-(1+2\beta)\ell j +2(r/s)nj -(r/s)j^{2}]},
\end{split}
\label{Eq:etainvariance1}
\end{equation}
where now we use Eq.\ (\ref{Eq:etainvariance1}) to describe the $n+1$th kick.  Noting that $e^{i\pi\ell j}=e^{-i\pi\ell j}$ and taking $n$ to be a multiple of $s$ (i.e., $n= \tau s$) yields
\begin{equation}
\tilde{\nu}_{\tau s+1}(\beta)
= e^{i\pi(1+2\beta)\ell s \tau -rs\tau^{2}}
\sum_{j=0}^{\tau s}
e^{i\pi[(1-2\beta)\ell j -(r/s)j^{2}]}.
\label{Eq:etainvariance2}
\end{equation}
Hence, as $\eta_{\tau s+1}(\beta) \equiv |\tilde{\nu}_{\tau s+1}(\beta)|$, it follows that $\eta_{\tau s+1}(\beta)=\eta_{\tau s+1}(-\beta)$ in general, whenever $\Omega=r/s$. If the initial momentum distribution is symmetric, all odd moments will be zero at these times; for integer $\Omega$, this means the odd moments will always be zero.

If $n=\tau s-1$, we deduce from Eqs.\ (\ref{Eq:etainvariance1}) and (\ref{Eq:etainvariance2}) that
\begin{equation}
\begin{split}
\tilde{\nu}_{\tau s}(\beta)
=& \sum_{j=0}^{\tau s}e^{i\pi[(1+2\beta)\ell j -(r/s)j^{2}]} - e^{i\pi(1+2\beta)\ell s\tau  -rs \tau^{2}}\\
=& e^{i\pi(1+2\beta)\ell s\tau -rs \tau^{2}}
\left\{
\sum_{j=0}^{\tau s}
e^{i\pi[(1-2\beta)\ell j -(r/s)j^{2}]} -1
\right\}.
\end{split}
\label{Eq:etainvariance3}
\end{equation}
Hence, it follows that $\eta_{\tau s}(\beta) = \eta_{\tau s}(-\beta)$ for the values of $\beta$ satisfying $(1+2\beta)\ell s \tau - rs\tau^{2} = 2A$ (i.e., for $\beta$ an even integer).  This condition is fulfilled whenever Eq.\ (\ref{rescondition}) or Eq.\ (\ref{antirescondition}) (for even $\tau$) hold, i.e., for values of $\beta$ where resonances and antiresonances are supported. 

Finally, if the initial momentum distribution is symmetric, all odd momentum moments will be zero when $n$ is a multiple of $s$, for resonant evolution, and when $n$ is an even multiple of $s$, for antiresonant evolution.  For this reason, we consider only even momentum moments in the remainder of this work.

\subsubsection{Time evolution of the even momentum moments}
The time evolution of the even momentum moments is given by
\begin{equation}
\begin{split}
\expec{\hat{P}^{2m}}_n=\int_{-1/2}^{1/2}d\beta \sum_{j,k=-\infty}^{\infty}J_j^2\left( \phi_d \eta\right)(j+k+\beta)^{2m}D_k(\beta),
\end{split}
\label{p2m}
\end{equation}
where we have shifted the index $j$ in Eq.\ (\ref{pm}) by $k$.  Note that [by Eq.\ (\ref{eta})] $\eta(\beta) = \eta(k+\beta)$.  Making the change of variables $P=k+\beta$ and binomially expanding $(j+P)^{2m}$ implies
\begin{equation}
\expec{\hat{P}^{2m}}_n
= \int_{-\infty}^{\infty}dP D(P) \sum_{j=-\infty}^{\infty} \sum_{h=0}^{m} {2m \choose 2h}J_j^2\left( \phi_d \eta\right)j^{2h}P^{2(m-h)}, \label{p2m0}
\end{equation}
where $D(P)=D_k(\beta)$ and we have eliminated odd powers of $j$, because they sum over $j$ to give zero [see Eq. (\ref{bes_sum_odd})].  Using the general form of the summation of Bessel functions over even powers of $j$ [see Eq. (\ref{bes_sum_ind})], we find that
\begin{equation}
\expec{\hat{P}^{2m}}_n= \expec{\hat{P}^{2m}}_0 +\int_{-\infty}^{\infty}dP D(P) \sum_{h=1}^{m}{2m \choose 2h}
R_{2h}(\phi_d\eta)
P^{2(m-h)}, 
\label{p2mofR}
\end{equation}
where $R_{2h}$ is a $2h$th degree even polynomial in $\phi_{d}\eta$.

Considering the two simplest even momentum moments [see Eq.\ (\ref{bes_sum_examples})], Eq.\ (\ref{p2mofR}) for $m=1$  becomes
\begin{equation}
\expec{\hat{P}^2}_n = \expec{P^2}_0 + \frac{\phi_d^2}{2} \int_{-\infty}^{\infty}dP  D(P)\eta^2, \label{p2master}
\end{equation}
and for $m=2$  becomes
\begin{equation}
\expec{\hat{P}^4}_n=\expec{P^4}_0 + \int_{-\infty}^{\infty}dP D(P)\left( \frac{3}{8}\phi_d^4\eta^4 + \frac{1}{2}\phi_{d}^{2}\eta^{2} + 3P^2\phi_d^2{\eta}^2\right). \label{p4master}
\end{equation}

\subsection{Ultracold limit \label{sec3c}}

\subsubsection{Time evolution of the moments at zero temperature}
\begin{figure}[tbp]
\centering
\includegraphics[width=8.8cm]{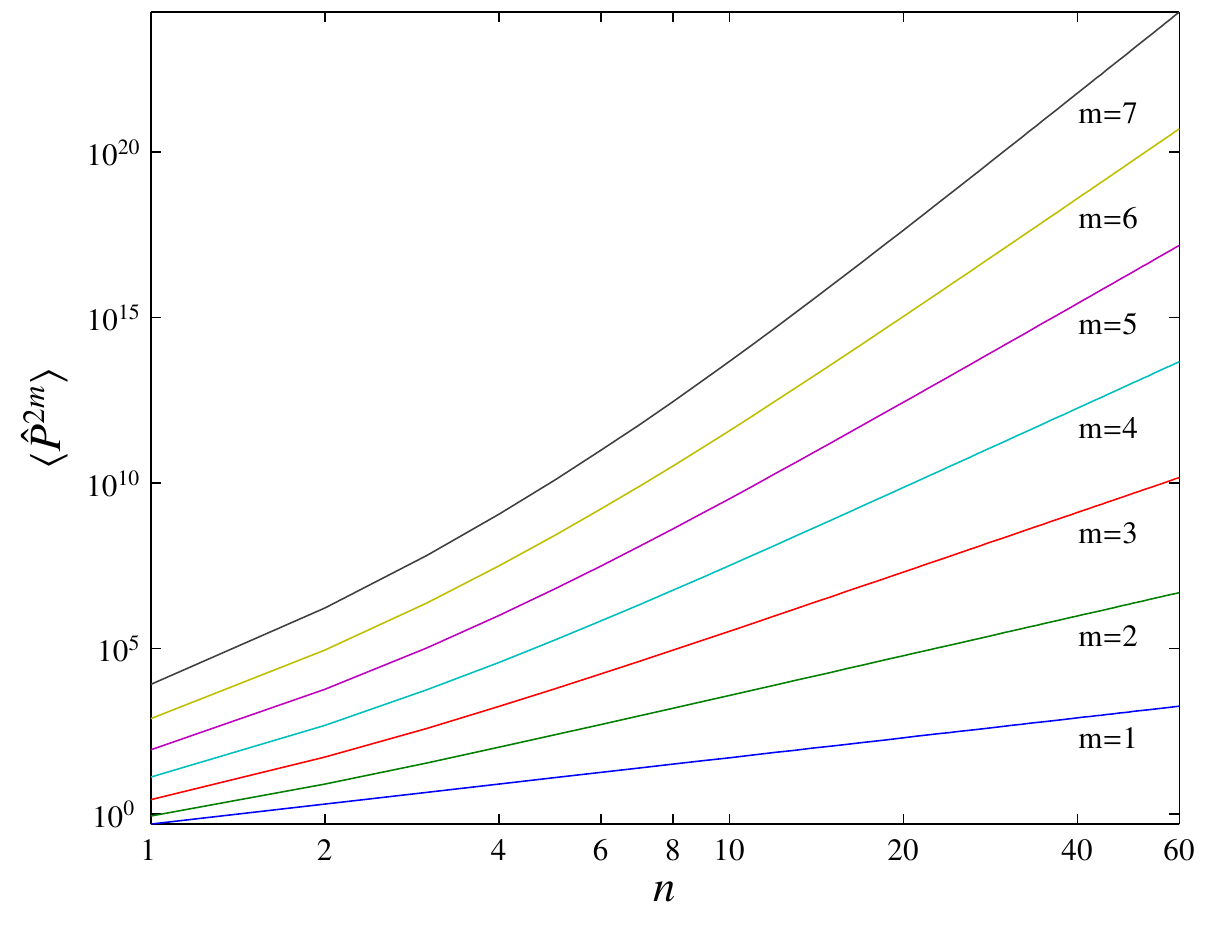}
\caption{(Color online)  The first seven even momentum moments of a $\delta$-kicked atomic cloud in the ultracold regime as given by Eq.\ (\ref{resp2mofR}) for integer $\Omega$, with the resonance condition (\ref{rescondition})  satisfied by choosing any appropriate value of $\ell$.  We choose $\phi_d=0.8\pi$ as corresponding to an experimentally typical driving strength \cite{dArcy2001a}.  The growth tends asymptotically to a power-law behavior with the power given by the order of the moment.}
\label{pmasnmfig}
\end{figure}

We consider the most trivial example of a symmetric initial momentum distribution, i.e., we choose $D(P)=\delta(P)$ such that all the atoms are initially in the $P=0$ state.  This describes an ideal zero-temperature gas, which we refer to as the ultracold limit.  In this case $\beta=0$; hence $\alpha=\ell$, and the resonance condition (\ref{rescondition}) becomes $s(r-\ell)=2A$.  Equation (\ref{p2mofR}) then simplifies to
\begin{equation}
\expec{\hat{P}^{2m}}_n= 
R_{2m}(\phi_{d}\eta),
\label{resp2mofR}
\end{equation}
i.e., the growth of $\expec{\hat{P}^{2m}}_n$ scales to leading order as $(\phi_{d}\eta)^{2m}$, as shown for integer $\Omega$ in Fig.\ \ref{pmasnmfig}.  In particular, we obtain
\begin{equation}
\expec{\hat{P}^2}_n = \frac{\phi_d^2}{2}\eta^2, 
\quad
\expec{\hat{P}^4}_n =\frac{3\phi_d^4}{8}\eta^4 + \frac{\phi_{d}^{2}}{2}\eta^{2},
\label{p2deltaeta}
\end{equation}
for $m=1$ and $m=2$, respectively.

\subsubsection{Integer values of $\Omega$}
The case where $\Omega = r/s$ is an integer is equivalent to choosing $s=1$.  The amplitude of Eq.\ (\ref{eta}) is then given by Eq.\ (\ref{omega_evaluate}) as
\begin{equation}
\eta = 
\left|\frac{\sin(n[\ell-r]\pi/2)}{\sin([\ell-r]\pi/2)}
\right|, 
\label{resmodetainteger}
\end{equation} 
which identifies $\eta$ as the modulus of Chebyshev's polynomial of the second kind \cite{Rivlin1974,Mason2003}, $U_{n-1}\left(\cos([\ell-r]\pi/2)\right)$.  Imposing the resonance condition $s(r-\ell)=2A$ with $s=1$ then gives $\eta=n$. 
We therefore deduce from Eq.\ (\ref{p2deltaeta}) that
\begin{equation}
\expec{\hat{P}^2}_n = \frac{ \phi_d^2}{2}n^2,\quad 
\expec{\hat{P}^4}_n = \frac{3\phi_d^4}{8}n^4 +  \frac{ \phi_d^2}{2}n^2. 
\label{resp2deltaint}
\end{equation}

Similarly, imposing the antiresonance condition (\ref{Nantirescondition}) implies $\ell - r = 2A + N_{r}/N_{s}$.  Hence, in the antiresonant case,
\begin{equation}
\eta = U_{n-1}\left((-1)^A\cos(\pi N_{r}/2N_{s})\right), 
\label{antiresmodetainteger}
\end{equation} 
and the antiresonant evolutions of the second- and fourth-order moments for integer $\Omega$ are given by
\begin{equation}
\begin{split}
\expec{\hat{P}^2}_n =& \expec{P^2}_0 + \frac{\phi_d^2}{2}U^2_{n-1}\left(\cos(\pi N_{r}/2N_{s})\right),
\\
\expec{\hat{P}^4}_n=& \expec{P^4}_0 + \frac{3\phi_d^4}{8}U^4_{n-1}\left(\cos(\pi N_{r}/2N_{s})\right) 
\\& + \frac{\phi_d^2}{2}U^2_{n-1}\left(\cos(\pi N_{r}/2N_{s})\right).
\end{split}
\label{antip2deltaint}
\end{equation} 

As an example, we consider the simplest antiresonance with $N_{r}=N_{s}=1$, so that $U_{n-1}=(-1)^n\sin\left(n\pi/2\right)$.  From Eq.\ (\ref{antip2deltaint}) we find that
\begin{equation}
\begin{split}
\expec{\hat{P}^2}_n =& \expec{P^2}_0 + \frac{\phi_d^2}{4}\left[(-1)^{n+1}+1\right],
\\
\expec{\hat{P}^4}_n =& \expec{P^4}_0 + 
\left(
\frac{3\phi_d^4}{16}
+ \frac{\phi_d^2}{4}\right)
\left[(-1)^{n+1}+1\right],
\end{split}
\label{antip2deltaintn1}
\end{equation}
demonstrating that the second- and fourth-order momentum moments oscillate with a period of two kicks.

\subsubsection{Rational values of $\Omega$} \label{Om_1overs}
If $\Omega=r/s$, then, for $n>s$, $\tilde{\nu}$ may be divided into identical summations with a remainder term.  Hence, assuming the resonance condition (\ref{rescondition}) to be fulfilled,
setting $n=\tau s + \lambda$ gives
\begin{equation}
\tilde{\nu} = \tau\sum_{j=0}^{s-1}e^{i\pi[\ell j - (r/s)j^2]} + \sum_{j=0}^{\lambda-1}e^{i\pi[\ell j-(r/s) j^2]}.
\label{etasplit}
\end{equation}
It is possible to rewrite the first summation using the reciprocity formula \cite{Berndt1981}:
\begin{equation}
\sum_{j=0}^{|C|-1}e^{i\pi[(A j^2+B j)/C]} = \sqrt{\left|\frac{C}{A}\right|}e^{i\pi[ (|A C|-B^2)/4AC]}\sum_{j=0}^{|A|-1}e^{-i\pi[(C j^2+B j)/A]}, 
\label{LS}
\end{equation} 
where $A$, $B$, and $C$ are integers such that $AC-B$ is even. This restriction is in fact exactly equivalent to the resonance condition (\ref{rescondition}).  Consequently,
\begin{equation}
\tilde{\nu} = \tau\sqrt{\frac{s}{r}}e^{i\pi(\ell^2 s/4r-rs)}\sum_{j=0}^{|r|-1}e^{i\pi [(sj^2+s\ell j)/r]} 
+ \sum_{j=0}^{\lambda-1}e^{i\pi[ \ell j-(r/s) j^2]}. 
\label{etaLS}
\end{equation} 
We can then impose $\lambda=0$ because, for large values of $\tau$ under conditions of quantum resonance, the second summation in Eq.\ (\ref{etaLS}) is only a small fluctuation in $\tilde{\nu}$.  Taking the simplest rational value of $\Omega=1/s$, Eq.\ (\ref{etaLS}) then simplifies considerably to give 
\begin{equation}
\eta = \tau \sqrt{s}. \label{etarat}
\end{equation} 
Substituting Eq.\ (\ref{etarat}) into Eq.\ (\ref{p2deltaeta}), and letting $\tau=n/s$,  yields
\begin{equation}
\expec{\hat{P}^2}_n = \frac{\phi_d^2}{2s}n^2, \quad
\expec{\hat{P}^4}_n = \frac{3\phi_d^4}{8s^2}n^4 + \frac{\phi_d^2}{2s}n^2,
\label{resp2deltarat}
\end{equation}
where it should be noted that these expressions are strictly valid only when $n$ is a multiple of $s$.  Hence,  in addition to the conventional quantum resonances associated with the $\delta$-kicked rotor/particle \cite{Saunders2007} (i.e., when $\Omega=0$) there are numerous \textit{fractional quantum resonances\/} observable in the $\delta$-kicked accelerator when $\Omega$ takes rational values.  The manifestation of such fractional quantum resonances in an atom-optical context at finite temperature we address in detail in a companion paper \cite{Saunders2008}.

\subsection{Finite-width momentum distributions \label{sec3d}}

\subsubsection{General Form of $\eta^{2q}$}
From Sec.\ \ref{sec3c} it is apparent that elucidation of the time evolution of the momentum moments is dependent on knowledge of the behavior of $\eta$.  It is not feasible to evaluate $\eta$ for all cases. However, it is possible to consider its general behavior within certain approximations.  

From the definition of $\eta$ [see Eq.\ (\ref{eta})], it follows that 
\begin{equation}
\begin{split}
\eta^{2q} = \left|\sum_{j=0}^{n-1}e^{i\pi(\alpha j-\Omega j^2)}\right|^{2q}
= \sum^{n-1}_{\mathbf{j}_q,\mathbf{j}^\prime_q=0}
e^{i\pi[F(\mathbf{j}_q)+\beta G(\mathbf{j}_q) - F(\mathbf{j}_q^\prime)-\beta G(\mathbf{j}_q^\prime)]},
\end{split}
\label{eta2q}
\end{equation}
where
\begin{subequations}
\begin{align}
F(\mathbf{j}_q) &= \sum_{r=1}^{q} (\ell j_r-\Omega j^2_r ) \label{FG1} \\ 
G(\mathbf{j}_q) &= 2\ell \sum_{r=1}^{q} j_r. \label{FG2}
\end{align}
\end{subequations}
For brevity, we have used the convenient shorthand $\sum_{\mathbf{j}_q,\mathbf{j}^\prime_q}$ to denote the $2q$ sums, with $\mathbf{j}_q=(j_1, \ldots,j_q)$ and $\mathbf{j}^\prime_q=(j^\prime_1, \ldots,j^\prime_q)$.  Substituting Eq.\ (\ref{eta2q}) into Eq.\ (\ref{p2mofR}) then yields
\begin{equation}
\begin{split}
\expec{\hat{P}^{2m}} =&   \expec{\hat{P}^{2m}_0} 
+  \sum_{h=1}^{m}{2m \choose 2h}
\sum_{q=1}^{h}a_{q}\phi_{d}^{2q}
\sum_{\mathbf{j}_q,\mathbf{j}_q^\prime=0}^{n-1} e^{i\pi [F(\mathbf{j}_{q})- F(\mathbf{j}_{q}^\prime)]} \\ &\times \int_{-\infty}^{\infty}dP D(P)P^{2(m-h)}e^{i\pi P [G(\mathbf{j}_{q})- G(\mathbf{j}_{q}^\prime)]},
\end{split}
\label{p2mbeforebroad}
\end{equation}
where we have 
expanded $R_{2h}(\phi_{d}\eta)=\sum_{q=1}^{h}a_{q}(\phi_{d}\eta)^{2q}$, and replaced $\beta$ by $k+\beta\equiv P$.  

\subsubsection{Large finite-width limit\label{Sec:LargeFiniteWidth}}
The integrands in Eq.\ (\ref{p2mbeforebroad}) are in general oscillatory and, by the method of stationary phase, under many circumstances give a negligible contribution to the momentum momentum evolution.  We first separate out the integrals with non-oscillatory integrands.  In particular, if $G(\mathbf{j}_{q})= G(\mathbf{j}_{q}^\prime)$, then  $F(\mathbf{j}_{q})- F(\mathbf{j}_{q}^\prime)=2\Omega\rho(\mathbf{j}_q,\mathbf{j}_q^\prime)$, where $\rho(\mathbf{j}_q,\mathbf{j}_q^\prime)$ is an integer given by\footnote{A simple example is for $q=2$ where we have the constraint $j_1+ j_2 = {j_1^\prime} + {j_2^\prime}$.  Considering $(j_1+j_2)^2=({j_1^\prime} + {j_2^\prime})^2$ leads to the conclusion that $j_1^2+j_2^2-{j_1^\prime}^2-{j_2^\prime}^2 = 2({j_1^\prime}{j_2^\prime}-j_1j_2)$.} 
\begin{equation}
\rho(\mathbf{j}_q,\mathbf{j}_q^\prime) = \sum_{x=1}^{q}\sum_{y=x+1}^q(j_{x}' j_{y}' - j_x j_y). 
\label{rho}
\end{equation}
Hence, partitioning Eq.\ (\ref{p2mbeforebroad}), and using Eq.\ (\ref{rho}), yields
\begin{equation}
\begin{split}
\expec{\hat{P}^{2m}}_n =&   \expec{\hat{P}^{2m}}_0 
+\sum_{h=1}^{m}{2m \choose 2h}\expec{P^{2(m-h)}}_0
\\&\times
\sum_{q=1}^{h}a_{q}\phi_{d}^{2q}\sum_{\substack{\mathbf{j}_q,\mathbf{j}_q^\prime \\ G(\mathbf{j}_q)= G(\mathbf{j}_{q}^\prime)}} 
e^{i2\pi\Omega\rho(\mathbf{j}_q,\mathbf{j}_q^\prime)}
\\&
+\sum_{h=1}^{m}{2m \choose 2h}
\sum_{q=1}^{h}a_{q}\phi_{d}^{2q}
\sum_{\substack{\mathbf{j}_q,\mathbf{j}_q^\prime \\ G(\mathbf{j}_q)\neq G(\mathbf{j}_{q}^\prime)}}  
e^{i\pi [F(\mathbf{j}_q)- F(\mathbf{j}_{q}^\prime)]}
\\ &\times \int_{-\infty}^{\infty}dP D(P) P^{2(m-h)}e^{i\pi P[G(\mathbf{j}_q)- G(\mathbf{j}_{q}^\prime)]}. \label{p2mbroad}
\end{split}
\end{equation}

In Eq.\ (\ref{p2mbroad}), the number of terms in the first summation over $\mathbf{j}_{q}$ and $\mathbf{j}_{q}'$ is equal to the number of different ways $G(\mathbf{j}_q)= G(\mathbf{j}_{q}^\prime)$ can be satisfied. This is given by a degree $2q-1$ polynomial in $n$, i.e.,
\begin{equation}
S_{2q-1}(n) = \sum_{r=1}^{2q-1} b_r n^r.
\label{Eq:S_polynomial}
\end{equation}
The coefficients $b_r$ can readily be computed for a given $q$, as outlined in Appendix \ref{appdiceprob}.  Note also that, for integer values of $\Omega$, $\exp({i2\pi\Omega\rho(\mathbf{j}_q,\mathbf{j}_q^\prime)})$ collapses to unity, independent of the value of $\rho(\mathbf{j}_q,\mathbf{j}_q^\prime)$.  Hence, for integer $\Omega$,
\begin{equation}
\begin{split}
\sum_{\substack{\mathbf{j}_q,\mathbf{j}_q^\prime \\ G(\mathbf{j}_q)= G(\mathbf{j}_{q}^\prime)}}
e^{i2\pi\Omega\rho(\mathbf{j}_q,\mathbf{j}_q^\prime)} = S_{2q-1}(n),
\label{sumwconstraint}
\end{split}
\end{equation}
and substituting Eq.\ (\ref{sumwconstraint}) into Eq.\ (\ref{p2mbroad}) yields
\begin{equation}
\begin{split}
\expec{\hat{P}^{2m}}_n =&   \expec{\hat{P}^{2m}}_0 
+\sum_{h=1}^{m}{2m \choose 2h}\expec{P^{2(m-h)}}_0
\sum_{q=1}^{h}a_{q}\phi_{d}^{2q}
S_{2q-1}(n)
\\& +\sum_{h=1}^{m}{2m \choose 2h}
\sum_{q=1}^{h}a_{q}\phi_{d}^{2q}
\sum_{\substack{\mathbf{j}_q,\mathbf{j}_q^\prime) 
\\ G(\mathbf{j}_q)\neq G(\mathbf{j}_{q}^\prime)}}  
e^{i\pi[F(\mathbf{j}_q)- F(\mathbf{j}_{q}^\prime)]}
\\ &\times \int_{-\infty}^{\infty}dP D(P)P^{2(m-h)}e^{i\pi P [G(\mathbf{j}_q)- G(\mathbf{j}_{q}^\prime)]}. \label{p2mbroad1}
\end{split}
\end{equation}

It is illustrative to rewrite
\begin{equation}
\begin{split}
e^{i\pi P[G(\mathbf{j}_q)- G(\mathbf{j}_{q}^\prime)]} 
= e^{i2\pi P\ell\sum_{r=1}^{q} [j_r - j^\prime_r]}, 
\label{sigmaexp}
\end{split}
\end{equation}
from which it is apparent that Eq.\ (\ref{sigmaexp}) has a maximum period in $P$ of $1/\ell$.  We can therefore consider initial distributions $D(P)$ with characteristic width $\gg 1/\ell$ to be ``broad,'' meaning that the oscillatory terms in the integrals of Eq.\ (\ref{p2mbroad1}) cause them to average essentially to zero.  This leaves
\begin{equation}
\expec{\hat{P}^{2m}}_n \approx \expec{\hat{P}^{2m}}_0 
+\sum_{h=1}^{m}{2m \choose 2h}\expec{P^{2(m-h)}}_0
\sum_{q=1}^{h}a_{q}\phi_{d}^{2q}
S_{2q-1}(n),\label{p2mbroad2}
\end{equation}
where the leading order is $n^{2m-1}$.  Therefore, for sufficiently broad distributions, we expect the $2m$th-order moment to scale with the number of kicks $n$ as $n^{2m-1}$ for integer values of $\Omega$.  

For $\Omega=r/s$, the $e^{i2\pi\Omega\rho(\mathbf{j}_{q},\mathbf{j}_{q}')}$ terms take $s$ different values.  If these values to some extent add in phase, then we expect the growth of the $2m$th-order moment to have the same power law as for integer values of $\Omega$, but with a constant coefficient that may depend on $\Omega$.  As described in a companion paper \cite{Saunders2008}, through a combination of numerical and analytical investigation we have found that, for a broad Gaussian initial momentum distribution, the growth of the second-order moment appears to be largely independent of the value of $\Omega$.  The growth of the fourth-order moment, however, is generally cubic with $n$ for rational $\Omega=r/s$, but with a constant coefficient that diminishes with increasing $s$.

\subsection{Cumulants} \label{sec3e}
The $2m$th-order moment also includes information regarding all lower-order moments, so observations of the moments alone do not explicitly isolate effects of different order.  However, the moments $\expec{\hat{P^m}}$ may be used to construct cumulants $\langle \langle \hat{P}^m\rangle \rangle$ \cite{Bach2005,Kubo1962,Fricke1996,Kohler2002,Gardiner1996}, which are independent quantities.  The first-, second-, third- and fourth-order cumulants are the mean, variance, skew, and kurtosis, respectively, and are given in terms of the moments by:
\begin{subequations}
\begin{align}
\langle \langle \hat{P}\rangle \rangle =& \expec{\hat{P}}, \\
\langle \langle \hat{P}^2\rangle \rangle =& \expec{\hat{P}^2}-\expec{\hat{P}}^2, \\
\langle \langle \hat{P}^3\rangle \rangle =& \expec{\hat{P}^3}-3\expec{\hat{P}}\expec{\hat{P}^2}+2\expec{\hat{P}}^3, \\
\langle \langle \hat{P}^4\rangle \rangle =& \expec{\hat{P}}^4 - 3\expec{\hat{P}^2}^2 + 12\expec{\hat{P}}^2\expec{\hat{P}^2} - 6\expec{\hat{P}}^4. 
\end{align}
\label{Eq:CumulantDefs}
\end{subequations}
The skew quantifies the asymmetry of the distribution about the mean, and the kurtosis quantifies the degree to which the distribution is peaked.  For example, a Gaussian (or $\delta$-function, which can be defined as a zero-variance limit of a Gaussian) has kurtosis $=0$, whereas a distribution which is more sharply peaked or cusp-like has positive kurtosis, and a distribution which is more ``blunt'' has negative kurtosis.  

As discussed in Sec.\ \ref{sec3b}, we are largely considering symmetric distributions, where all odd moments, and therefore  cumulants, are zero.  In this instance, the two lowest-order non-zero cumulants are given by:
\begin{equation}
\langle \langle \hat{P}^2\rangle \rangle = \expec{\hat{P}^2},
\quad
\langle \langle \hat{P}^4\rangle \rangle = \expec{\hat{P}}^4 - 3\expec{\hat{P}^2}^2.
\label{cumform}
\end{equation}
The evolutions of the second- and fourth-order moments in the ultracold regime [$D(P)=\delta(P)$] for $\Omega=r/s$ are described by Eq.\ (\ref{resp2deltarat}). From this the lowest even cumulants, as defined in Eq.\ (\ref{cumform}), readily follow:
\begin{equation}
\langle \langle \hat{P}^2\rangle\rangle_n = \frac{\phi_d^2}{2s}n^2,  \quad
\langle \langle \hat{P}^4\rangle\rangle_n = \frac{\phi_d^2}{2s}n^2 - \frac{3\phi_d^4}{8s^2}n^4. 
\label{asymnarrow2}
\end{equation}
Hence, the leading-order power-law behaviour manifest in the lowest two even moments is also observed in the lowest two even cumulants.  We have investigated numerically the behaviour of all even cumulants up to $m=100$, which also exhibit the same leading $m$th-order power-law behaviour as the $m$th moments.  The scaling laws therefore do not appear to be simply manifestations of lower-order effects.

Although the cumulant definitions of Eq.\ (\ref{Eq:CumulantDefs}) are formally convenient, determining the moments from an experimentally measured momentum distribution and then combining them into cumulants will cause experimental errors to accumulate.  Alternative, more direct assessment of the ``width'' and ``pointedness'' of the measured distribution, providing it yields essentially the same information on the system dynamics,  may turn out to be experimentally more convenient.

\section{Finite temperature ideal gas} \label{sec4}
\subsection{Overview}

In Sec.\ \ref{Sec:LargeFiniteWidth} we determined that for an initial momentum distribution that is symmetric and in some sense sufficiently broad, the $2m$th-order momentum moments are expected to grow as $n^{2m-1}$, where $n$ is the kick number.  That is, the power-law growth is in general reduced by one compared to the ultracold limit (Sec.\ \ref{sec3c}).  We can illustrate this with analytical expressions for  the second- and fourth-order moments and cumulants when the initial momentum distribution is that of an ideal gas in thermal equilibrium, i.e., 
\begin{equation}
D(P)=\frac{1}{\sqrt{2\pi w^2}}\exp(-P^2/2w^2).
\label{Eq:GaussDist}
\end{equation} 
Here $w$ is the standard deviation and the corresponding Boltzmann temperature is given by $\mathcal{T}_{w}=\hbar^{2}K^{2}w^{2}/Mk_{B}$, where $k_{B}$ is Boltzmann's constant.  

Here we choose to constrain $\Omega$ to take integer values $r$, that is we do not consider the fractional quantum resonances derived for $\Omega=1/s$ in the ultracold limit in Sec.\ \ref{Om_1overs}.  We note, however, that when considering the manifestation of fractional quantum resonant dynamics in a finite temperature ideal gas, the behaviour of the fourth moment distinguishes clearly between different kinds of fractional quantum resonances, whereas the behaviour of the second moment does not \cite{Saunders2008}.  This provides an additional, specific motivation for better understanding the dynamics of the fourth-order momentum moment.

When comparing with experiment, we note that our analytical results implicitly assume that the momentum distribution can be determined with perfect precision.  Obviously this cannot be exactly fulfilled experimentally (although Gustavsson \textit{et al}.\  \cite{Gustavsson2008b} report measuring the momentum distribution of cold caesium atoms with a resolution equivalent to $0.05\hbar K$).\footnote{Note that $\hbar K$ is equal to two photon recoils.} Our numerical investigations indicate that, at least for the second- and fourth-order moments, the qualitative behaviour of the moments is not very sensitive to coarsening of the momentum resolution.  Also note that, as quantum resonant dynamics are associated with the momentum of a portion of the atomic cloud increasing ballistically,  as time progresses a proportion of the atoms will no longer be in the Raman-Nath regime.  Experimentally, we therefore expect the predicted power-law behaviour to be followed for a finite time only, as dictated by details of the experimental configuration.

\subsection{Calculation of the Moments}
\label{gaussdist}

\begin{figure}[tbp]
\centering
\includegraphics[width=8.8cm]{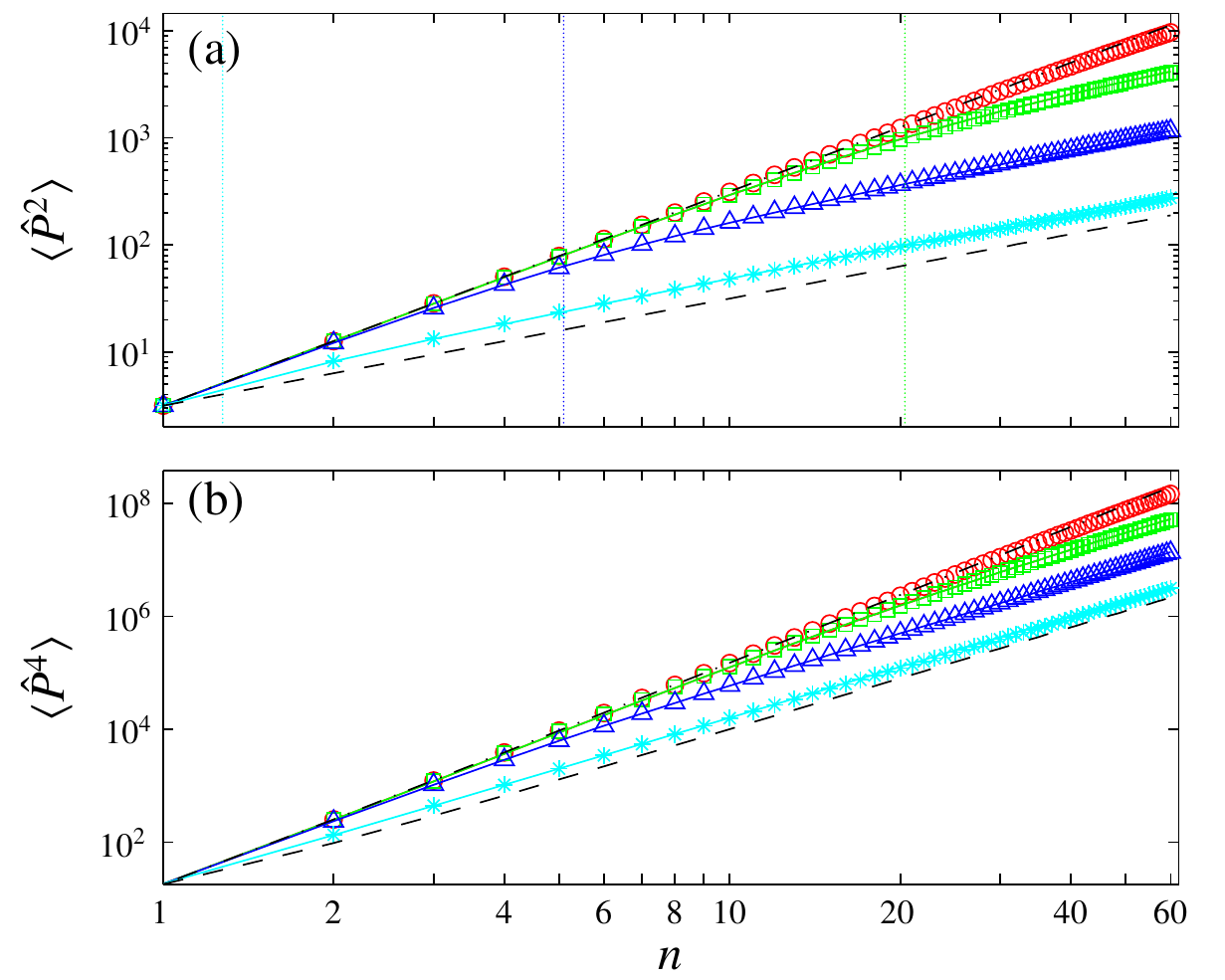}
\caption{(Color online)  
Second-order (a) and fourth-order (b) momentum moments yielded by the quantum-resonant evolution of an atomic cloud, as propagated by Eq.\ (\ref{F}).  We set $\Omega=0$ (i.e., no net acceleration), $T=T_{T}$ ($\ell=2$), and choose $\phi_{d}=0.8\pi$ \cite{dArcy2001a}.  The initial momentum distributions are Gaussian [see Eq.\ (\ref{Eq:GaussDist})] with standard deviations $w$ given by $(\textcolor{cyan}{\ast})$ $w=1/8$, $(\textcolor{blue}{\triangle})$ $w=1/32$, $(\textcolor{green}{\Box})$ $w=1/128$, $(\textcolor{red}{\circ})$ $w=1/512$. The solid lines correspond to analytic results given by Eqs.\ (\ref{p2_gauss}) and (\ref{p4_gauss}), and the symbols correspond to data from a Monte Carlo simulation, the details of which are discussed elsewhere\cite{Saunders2007}.  The dashed-dotted lines indicate the ultracold limit [see Eq.\ (\ref{asymnarrow2}) with $s=1$], and the dashed lines are a lower bound limit for a broad distribution [see Eq.\ (\ref{gauasymbroad}) with $w$ set $=0$].  The vertical dotted lines in (a) indicate $n_{G}= 1/\ell\pi w$ for $w=1/8,1/32,1/128$, i.e., the discrete time characterizing when growth in $\expec{\hat{P}^{2}}$ deviates from quadratic to linear.}
\label{figbro}
\end{figure}

The general procedure is outlined in Appendix \ref{appdist}; starting with Eqs.\ (\ref{p2master}) and (\ref{p4master}), we evaluate $\eta$, expand the resulting expression in terms of cosines, and then integrate. Using Eqs.\ (\ref{p2master}) and (\ref{p4master}) to determine Eqs.\ (\ref{Eq:AppGauss2}) and (\ref{appp4_gauss_res}), we find that
\begin{equation}
\begin{split}
\expec{\hat{P}^2}_n =& w^2 + \frac{\phi_d^2}{2}n + \phi_d^2\sum_{q=1}^{n-1}(-1)^{q(\ell-r)}(n-q)e^{-2q^2\ell^2\pi^2w^2}, \label{p2_gauss}
\end{split}
\end{equation}
and
\begin{equation}
\begin{split}
\expec{\hat{P}^4}=& 3w^4  + \frac{\phi_d^4}{8}(2n^2+1)n+\frac{\phi_d^2}{2}n  + 3w^2\phi_d^2n
\\&+ \frac{3\phi_d^4}{8}\sum_{q=1}^{n-1}(-1)^{q(\ell-r)}e^{-2q^2\ell^2\pi^2w^2}
\\ &\times \left[q^3 -2nq^2-q +\frac{2n}{3}(2n^2+1)\right]\\&-  \frac{3\phi_d^4}{8}\sum_{q=n}^{2n-2}(-1)^{q(\ell-r)}e^{-2q^2\ell^2\pi^2w^2} 
\\ &\times \left[ \frac{q^3}{3}-2nq^2+\frac{(12n^2-1)}{3}q +\frac{2n}{3}(1-4n^2)\right]
\\&+\phi_d^2\sum_{q=1}^{n-1}(-1)^{q(\ell-r)}(n-q)e^{-2q^2\ell^2\pi^2w^2}
\\&+6\phi_d^2\sum_{q=1}^{n-1}(-1)^{q(\ell-r)}(n-q)w^2(1-4w^2q^2\ell^2\pi^2) 
\\ &\times e^{-2q^2\ell^2\pi^2w^2}. \end{split}
\label{p4_gauss}
\end{equation}
The timescale over which the resonant growth of the second-order moment (where $\ell-\Omega$ is even) switches from the ultracold regime to the large finite-temperature regime can be deduced from
Eq.\ (\ref{p2_gauss}).  A substantially diminished growth of the second-order moment occurs when $q^2\ell^2\pi^2w^2 \gg 1$, so we take $n_G=1/\ell\pi w$ to be the transition time.  Assuming that $w$ is sufficiently large that $n_G < 1$, the exponential terms in Eqs.\ (\ref{p2_gauss}) and (\ref{p4_gauss}) become vanishingly small, and we find that
\begin{subequations}
\begin{align}
\expec{\hat{P}^2}_n =& w^2 + \frac{\phi_d^2}{2}n, \label{gauasymbroad2}\\
\expec{\hat{P}^4}_n=& 3w^4 + \frac{\phi_d^4}{8}(2n^2+1)n +\frac{\phi_d^2}{2}(6w^2 + 1)n. \label{gauasymbroad4}
\end{align}
\label{gauasymbroad}
\end{subequations}
The transition time characterizing the transition from quartic to cubic growth of the fourth-order moment is not straightforward to define.  However, from Eq.\ (\ref{p4_gauss}) and Fig.\ \ref{figbro}, it is clear that this occurs over the same timescale as for the second-order moment.  In previous work  we considered a symmetric initial momentum distribution that is uniform over a finite range (less than $1/\ell$) \cite{Saunders2007}; the same qualitative behaviour is then manifest as in Fig.\ (\ref{figbro}), with comparable timescales.

\subsection{Momentum cumulants and the thermal limit \label{sec4c}}
\begin{figure}[tbp]
\centering
\includegraphics[width=8.8cm]{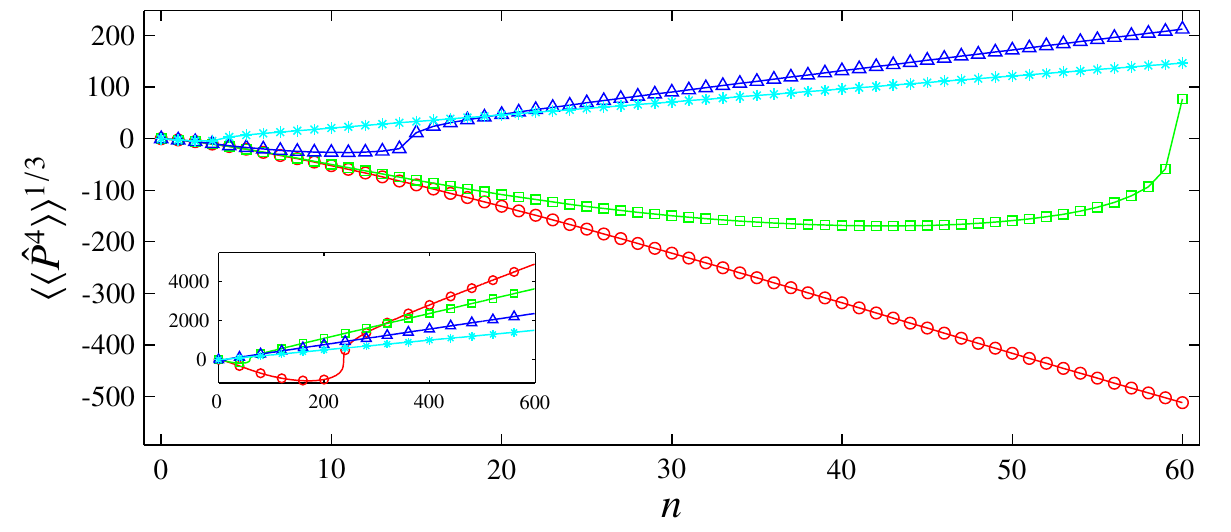}
\caption{(Color online) Cube-roots of the fourth-order momentum cumulants of a resonantly kicked atomic cloud evolved according to Eq.\ (\ref{F}).  Parameters are $\Omega=0$, $T=T_{T}$ ($\ell=2$), and we choose $\phi_{d}=0.8\pi$~\cite{dArcy2001a}.
We consider Gaussian initial momentum distributions with standard deviation $(\textcolor{cyan}{\ast})$ $w=1/8$, $(\textcolor{blue}{\triangle})$ $w=1/32$, $(\textcolor{green}{\Box})$ $w=1/128$, $(\textcolor{red}{\circ})$ $w=1/512$. After a certain number of kicks, the growth becomes linear, i.e., the fourth-order cumulant grows cubically. The inset shows the ideal (assuming perfect $\delta$-kicks) long-time behavior.}
\label{figcomp}
\end{figure}

As discussed in Sec.\ref{sec3e}, the momentum moments are in general not independent quantities and include lower-order correlations.  It is therefore instructive to consider the second- and fourth-order cumulants derived from a Gaussian initial momentum distribution [Eq.\ (\ref{Eq:GaussDist})].
From Eqs.\ (\ref{cumform}) and (\ref{gauasymbroad}) it follows that $\langle\langle \hat{P}^{2}\rangle\rangle_{n}= \langle \hat{P}^{2}\rangle_{n}$ and
\begin{equation}
\langle \expec{\hat{P}^4}\rangle_n= \frac{\phi_d^4}{8}(2n^2-6n+1)n + \frac{\phi_d^2n}{2}. 
\label{p4_gauss_vbroad}
\end{equation}
Hence, as in the ultracold limit [Eq.\ (\ref{asymnarrow2})], the power laws derived for the momentum moments in the thermal regime also appear to hold for the cumulants; the long-term cubic power-law growth of $\langle \expec{\hat{P}^4}\rangle_n$ is illustrated in Fig.\ \ref{figcomp}.  As in the ultracold limit (Sec.\ \ref{sec3e}), the scaling laws are therefore not simply manifestations of lower-order effects.  Although we have only shown this for the second- and fourth-order cumulants, we expect it to be generally true.  

Note also that both the moments and the \textit{change\/} in the fourth-order moment (i.e., $\langle\hat{P}^{4}\rangle_{n} - \langle\hat{P}^{4}\rangle_{0}$) formally diverge for large $w$, whereas this is not the case for the changes in the cumulants from their initial values.  It is therefore in the cumulants, rather than the moments, that one can speak of a well-defined thermal limit \cite{Saunders2008}.

\section{Conclusions\label{sec5}}

We have studied the quantum-resonant dynamics of the atom-optical quantum $\delta$-kicked accelerator, a fundamental system in the study of quantum chaos which reduces to the $\delta$-kicked particle (the atom-optical realization of the quantum $\delta$-kicked rotor) when the (tunable) effective acceleration parameter $a=0$.  In the $\delta$-kicked particle, quantum resonant antiresonant dynamics result when the periodicity of repeated off-resonant laser pulses is a half-integer multiple of the the Talbot time $T_{T}=2\pi M/\hbar K^{2}$.  We have found that, for such pulse periodicities, fractional quantum resonances occur when the rescaled effective acceleration parameter $\Omega = a K T^{2}/2\pi$ takes non-integer rational values.  

We have focussed on the dynamics of the momentum moments of the atomic centre-of-mass momentum distribution.  The moment evolutions are useful for characterizing the different kinds of dynamic behavior in the atom-optical $\delta$-kicked accelerator.  We have found explicitly that, in the ultracold limit, quantum resonant dynamics cause all even-ordered momentum moments to exhibit a power-law growth, with an exponent equal to the order of the moment.  For fractional quantum resonances the moment growth follows the same power laws, but with coefficients that reduce with increasing size of the denominator of $\Omega$.  We have argued that, for a sufficiently broad initial atomic centre-of-mass momentum distribution undergoing quantum resonant, or fractional quantum resonant evolution, the exponent of the power law growth of the even-ordered moments  will generally be reduced by one.  In every case considered, we have found that the long-term momentum cumulant dynamics also display power-law growth with the same exponent, meaning the scaling laws in the growth of the moments are not simply manifestations of lower-order effects.

We have illustrated a reduction in the power-law exponent by calculating explicit expressions for the dynamics of the second- and fourth-order momentum moments for a finite timperature ideal gas initially at thermal equilibrium, i.e., using a Maxwell-Boltzmann distribution to describe the initial state.  These are the two simplest moments to calculate, and are  physically motivated in that they provide directly useful information on the dynamical behavior of the atomic cloud: the second-order moment provides a signature to distinguish between dynamical localization and quantum-resonant dynamics in the atom-optical $\delta$-kicked particle \cite{Saunders2007} and accelerator, and the fourth-order momentum moment distinguishes between different classes of fractional resonant dynamics in the quantum $\delta$-kicked accelerator \cite{Saunders2008}.


\section*{Acknowledgements}
We would like to thank I. G. Hughes,  M. Edwards, K. Helmerson, and W. D. Phillips for useful and stimulating discussions, and V. J. Armitage for discussions regarding Gauss sums. We acknowledge funding from the UK EPSRC (Grant No. EP/D032970/1) and P.L.H. also thanks Durham University.

\begin{appendix}

\section{Momentum-state time evolution \label{matrix_element}}
\subsection{Overview}
In this appendix we determine the time evolution of the eigenstate $\ket{k+\beta}$ by applying the Floquet operator (\ref{F}) consecutively for $n$ kicks.  Our derivation follows closely the equivalent calculation for the $\delta$-kicked rotor (where $\Omega=0$) \cite{Saunders2007}.
\subsection{Application of the transformed Floquet operators}
The effect of applying the $n$th transformed Floquet operator (\ref{F}) to the momentum eigenstate $\ket{k+\beta}$ is
\begin{equation}
\begin{split}
\tilde{F}_n (\beta) \ket{k+\beta}  = &\int dz \tilde{F}_n (\beta) \ket{z} \braket{z}{k+\beta}  \\  
= & e^{ i[ \pi (1+\beta) \ell -K \gamma_n] \beta} 
\\ & \times
\int dze^{ -i \hat{k} K \gamma_n } e^{ i\phi_{d} \cos(K\hat{z}) } \ket{z} \braket{z}{k+\beta} ,\end{split}
 \label{OneKick}
\end{equation}
where we have defined
\begin{equation}
K \gamma_{n} \equiv \pi \left[ (1+2\beta)\ell - \Omega\left(2n-1\right)\right].
\label{Eq:Kgamma_n}
\end{equation}
Using $\exp(-i\hat{k}K\gamma_n)\ket{z}=\exp(i\beta K \gamma_n)$, and $\braket{z}{k+\beta}=\sqrt{K/2\pi}\exp[i(k+\beta)Kz]$, we deduce from Eq.\ (\ref{OneKick}) that
\begin{equation}
\begin{split}
\hat{F}_n (\beta) |k+\beta\rangle =& \sqrt{\frac{K}{2\pi}} 
e^{i\pi\beta(1+\beta)\ell} \int dz|z+\gamma_n \rangle e^{i(k+\beta)Kz} \\ &\times e^{i\phi_{d}\cos(Kz)},
\label{one_kick}
\end{split}
\end{equation}
where we have used $\langle z|k+\beta\rangle = \sqrt{K / 2\pi} e^{ i(k+\beta)Kz}$.

The combined effect of the $n$th and $(n+1)$th transformed Floquet operators can be determined similarly to obtain
\begin{equation}
\begin{split}
\tilde{F}_{n+1} (\beta) \tilde{F}_n (\beta) |k+\beta \rangle
= & \sqrt{\frac{K}{2\pi}} 
e^{i2\pi\beta(1+\beta)\ell} 
\int dz |z+\gamma_n+\gamma_{n+1} \rangle  \\
& \times e^{i(k+\beta)Kz}
e^{ i\phi_{d} [ \cos(Kz)+\cos(Kz+K\gamma_n) ]}.
\end{split}
\end{equation}
Hence, the full time evolution of the initial momentum eigenstate, governed by $\tilde{\mathcal{F}}_n (\beta)= \tilde{F}_n(\beta)\tilde{F}_{n-1}(\beta)\ldots\tilde{F}_1(\beta)$, is
\begin{equation}
\begin{split}
\tilde{\mathcal{F}}_n (\beta) |k+\beta \rangle
  = & \sqrt{\frac{K}{2\pi}} 
e^{in\pi\beta(1+\beta)\ell} 
  \int dz 
  \left| z +\sum_{n'=1}^{n}\gamma_{n'} \right\rangle
 \\
& \times    e^{ i(k+\beta)Kz}\exp \Biggl( i\phi_{d} \Biggl[ \cos(Kz) 
\\& 
+ \sum_{j=1}^{n-1}\cos\Biggl( K
z+\sum_{j'=1}^{j}K\gamma_{j'} \Biggr) \Biggr] \Biggr). 
\end{split}
\label{iteration}
\end{equation}

\subsection{Spatial representation}
To determine the matrix element $\protect{\bra{z}\tilde{\mathcal{F}}_n(\beta)\ket{k+\beta}}$, we project Eq.\ (\ref{iteration}) onto $\ket{z}$, i.e.,
\begin{equation}
\begin{split}
\langle z |\tilde{\mathcal{F}}_n (\beta) |k+\beta\rangle = 
& \sqrt{\frac{K}{2\pi}}
e^{in\pi\beta(1+\beta)\ell} 
 e^{  i(k+\beta)\left(Kz-\sum_{n'=1}^{n}K\gamma_{n'} \right) } \\
& \times \exp\left( i\phi_{d} \left[ \cos\left( Kz-\sum_{n'=1}^{n}K\gamma_{n'} \right) \right.\right.\\
&+ \left.\left.
\sum_{j=1}^{n-1}\cos\left( Kz-\sum_{n'=1}^{n}K\gamma_{n'}
+\sum_{j'=1}^{j}K\gamma_{j'} \right) \right]\right) . 
\label{difficult_sums}
\end{split}
\end{equation}
Using Eq.\ (\ref{Eq:Kgamma_n}), we can readily evaluate the sum
\begin{equation}
\sum_{n'=1}^{n} K\gamma_{n'}
= n \pi \left[(1+2\beta)\ell  -\Omega n\right] \label{Gsum},
\end{equation}
and, defining 
\begin{equation}
q_j\equiv (n-j)\pi \left[(1+2\beta)\ell-\Omega(n+j)\right], 
\label{q_j}
\end{equation}
Eq.\ (\ref{difficult_sums}) simplifies to
\begin{equation}
\begin{split}
\langle z |\tilde{\mathcal{F}}_n (\beta)|k+\beta\rangle = 
&\sqrt{\frac{K}{2\pi}}
e^{in\pi\beta(1+\beta)\ell} 
 e^{  i(k+\beta)\left(Kz-q_0\right)} \\
&\times\exp\left( i\phi_{d} \sum_{j=0}^{n-1}\cos( Kz-q_j)\right).
\label{eqn_half_way}
\end{split}
\end{equation}
Finally, defining $\xi \equiv \sum_{j=0}^{n-1}
\cos q_{j}$ and $\zeta  \equiv \sum_{j=0}^{n-1} \sin q_{j}$, Eq.\ (\ref{eqn_half_way}) further simplifies to
\begin{equation}
\begin{split}
\langle z |\tilde{\mathcal{F}}_n (\beta) |k+\beta\rangle = &
\sqrt{\frac{K}{2\pi}}
e^{in\pi\beta(1+\beta)\ell} 
 e^{  i(k+\beta)\left(Kz-q_0\right) } \\
& \times e^{i\phi_{d} \xi\cos(Kz)
+i\phi_{d}\zeta  \sin(Kz)}. \label{manipulate}
\end{split}
\end{equation}

\subsection{Probability amplitudes}

\subsubsection{General form of the probability amplitudes}

Invoking Bessel function expansions \cite{Saunders2007}, Eq.\ (\ref{manipulate}) can be recast in the form
\begin{equation}
\begin{split}
\langle z |\tilde{\mathcal{F}}_n (\beta)|k+\beta\rangle = &
\sqrt{\frac{K}{2\pi}}
e^{in\pi\beta(1+\beta)\ell} e^{  i(k+\beta)(Kz-q_0) }
\\ &\times 
\sum_{j=-\infty}^{\infty} e^{ijKz}J_{j}(\omega)e^{ij\chi},
\label{Exponential_term}
\end{split}
\end{equation}
where  $\omega$ and $\chi$ are real and defined by $\omega e^{i\chi} \equiv \phi_d ( i\xi+\zeta )$. To evaluate the matrix element
\begin{equation}
\begin{split}
 \bra{j+\beta^\prime} \tilde{\mathcal{F}}_n (\beta)\ket{k+\beta} 
&=\sqrt{\frac{K}{2\pi}}\int dz e^{-i(j+\beta^\prime)Kz}\bra{z}\tilde{F}_n(\beta)\ket{k+\beta}, 
\label{whyweneedspatial}
\end{split}
\end{equation}
we substitute Eq.\ (\ref{Exponential_term}) into Eq.\ (\ref{whyweneedspatial}) to yield
\begin{equation}
\begin{split}
\langle j+\beta'|\mathcal{F}_n (\beta))|k+\beta\rangle  = &  
e^{-i\pi n \alpha k}
e^{-in^{2}\pi(k+\beta)\Omega}
e^{-in\pi\beta^{2}\ell}
\sum_{j'=-\infty}^{\infty} J_{j'} ( \omega ) 
\\ & \times 
e^{ij'\chi}
\frac{K}{2\pi}\int dz e^{i(k+j'-j+\beta-\beta')Kz},
\end{split}
\label{ckj_one}
\end{equation}
where $\alpha = (1+2\beta)\ell$.

We now consider the general expansion
\begin{equation}
\ket{\Psi(t=nT)} \equiv \tilde{\mathcal{F}}_n (\beta)\ket{k+\beta} 
= \sum_{j=-\infty}^{\infty}c_{kj}(\beta,nT)\ket{j+\beta}, \label{timeevol}
\end{equation}
where the probability amplitudes $c_{kj}(\beta,nT)$ are given by
$c_{kj}(\beta,nT)\delta(\beta-\beta^\prime) = \langle j+\beta^\prime | \hat{\cal F}_n (\beta)|k+\beta \rangle$.  Evaluating the Fourier integral in Eq.\ (\ref{ckj_one}), it follows that
\begin{equation}
c_{kj}(\beta,nT) = 
J_{j-k} ( \omega ) 
e^{i(j-k)\chi}
e^{-i\pi n \alpha k}
e^{-in^{2}\pi(k+\beta)\Omega}
e^{-in\pi\beta^{2}\ell},
\label{Eq:p_amplitude}
\end{equation}
where the normalisation condition $\sum_{j=\infty}^{\infty}\left|c_{kj}(\beta,nT) \right|^2 = 1$ is satisfied.  Equations (\ref{timeevol}) and (\ref{Eq:p_amplitude}) describe the time evolution of the initial momentum eigenstate $\ket{k+\beta}$.  However, it still remains to evaluate $\omega$ and $\chi$.

\subsubsection{Evaluation of $\omega$ and $\chi$}
It is convenient to define\footnote{The parameter $\nu$ in this paper is related to a previously defined parameter $\mu$ \cite{Saunders2007} according to $\nu = i\mu^{*}$.} 
\begin{equation}
\phi_{d}\nu = \omega e^{i\chi},
\label{phidnu}
\end{equation}
i.e.,
\begin{equation}
\nu \equiv i\xi + \zeta = i \sum_{j=0}^{n-1} e^{-iq_j}.
\label{mu_eqn}
\end{equation}
Substituting Eq.\ (\ref{q_j}) into Eq.\ (\ref{mu_eqn}), we obtain
\begin{equation}
\nu = i e^{-i\pi(\alpha n-\Omega n^{2})}
\sum_{j=0}^{n-1} e^{i\pi(\alpha j - \Omega j^2)}.
\label{mu_simple}
\end{equation}
Absorbing the global phase, we define 
\begin{equation}
\tilde{\nu} = \sum_{j=0}^{n-1} e^{i\pi(\alpha j - \Omega j^2)},
\label{eta_def}
\end{equation}
and hence the magnitude $\eta\equiv |\tilde{\nu}| = |\nu|=|\omega|/\phi_{d}$.  It is perhaps natural to think of $\omega$ in $\omega e^{i\chi}$ as being positive, although we find it more convenient in Eq.\ (\ref{omega_evaluate}) to allow $\omega$ to take negative values.  

\subsubsection{Integer values of $\Omega$}
\label{appsin}

The Gauss sum in Eq.\ (\ref{eta_def}) can be evaluated analytically in some particular cases.  Here we illustrate this for integer $\Omega=r$.  For this choice of $\Omega$, Eq.\ (\ref{eta_def})  reduces to a geometric sum,
\begin{equation}
\tilde{\nu} = \sum_{j=0}^{n-1}e^{i\pi(\alpha-r)j},
\end{equation}
which can be evaluated to give
\begin{equation}
\tilde{\nu} = \frac{1-e^{in(\alpha-r)\pi}}{1-e^{i(\alpha-r)\pi}}
= e^{i(n-1)(\alpha-r)\pi/2}\frac{\sin(n[\alpha-r]\pi/2)}{\sin([\alpha-r]\pi/2)}.
\end{equation}
Referring to Eqs.\  (\ref{mu_eqn}) and (\ref{mu_simple}), we can now set
\begin{equation}
\frac{\omega}{\phi_{d}}= \frac{\sin(n[\alpha-r]\pi/2)}{\sin([\alpha-r]\pi/2)},
\label{omega_evaluate}
\end{equation}
and $\chi = [1-(\alpha -r)(n+1)]\pi/2$.  Note that Eq.\ (\ref{omega_evaluate}) can take negative values, and hence the absolute value must be taken to obtain $\eta$, if desired.

\section{Summations of Bessel functions\label{appbessum}}

We first state the recurrence relation \cite{Abramowitz1964}
\begin{equation}
J_{n-1}(x) + J_{n+1}(x) = \frac{2n}{x}J_n(x), 
\label{prop1}
\end{equation}
and Neumann's addition theorem \cite{Abramowitz1964}
\begin{equation}
\sum_{n=-\infty}^{\infty}J_{n}(x_1)J_{n+k}(x_2) = J_k(x_1-x_2), \label{prop2}
\end{equation}
where $n$ and $k$ are integer.  The addition theorem (\ref{prop2}) has two special cases:
\begin{equation}
\sum_{n=-\infty}^{\infty}J_n^2(x) = \sum_{n=-\infty}^{\infty}J_{n+k}^2(x) = 1 \label{Neum1}
\end{equation}
for any integer value of $k$, and,
\begin{equation}
\sum_{n=-\infty}^{\infty}J_n(x)J_{n+k}(x) = 0,
\label{Neum2}
\end{equation}
for integer $k\neq 0$.

The identity $J_{-n}(x) = (-1)^{n}J_{n}(x)$ implies  $J_{n}^{2}(x) = J_{-n}^{2}(x)$ \cite{Abramowitz1964}, and therefore that
\begin{equation}
\sum_{n=-\infty}^{\infty}n^{2m-1}J_{n}^{2} (x) = 0,
\label{bes_sum_odd}
\end{equation} 
for integer $m>0$.  In contrast, repeated substitution of the recurrence relation (\ref{prop1}) reveals
\begin{equation}
\sum_{n=-\infty}^{\infty}n^{2m}J_{n}^{2}(x) = 
\sum_{q=1}^{2m}
\left(\frac{x}{2}\right)^{q}
\sum_{p=0}^{q}
c_{q,p}
\sum_{n=-\infty}^{\infty}
J_{n-q+2p}(x)
J_{n}(x),
\label{Eq:BesselRecur}
\end{equation}
where it is in principle possible, although often tedious, to determine the  coefficients $c_{q,p}$. With Eqs.\ (\ref{Neum1}) and (\ref{Neum2}) we can eliminate all terms in Eq.\ (\ref{Eq:BesselRecur}) except those where $q=2p$.  Hence, 
\begin{equation}
\sum_{n=-\infty}^{\infty}n^{2m}J_{n}^{2} (x) =
\sum_{q=1}^{m}
a_{q}
x^{2q} \equiv R_{2m}(x),
\label{bes_sum_ind}
\end{equation}
where $a_{q}\equiv c_{2q,q}/2^{2q}$, and we note that the leading-order coefficient of the polynomial $R_{2m}(x)$ is always $a_{m} = {2m \choose m}/2^{2m}$.  In particular:
\begin{equation}
\sum_{n=-\infty}^{\infty}n^{2}J_{n}^{2} (x) =\frac{x^{2}}{2},
\quad
\sum_{n=-\infty}^{\infty}n^{4}J_{n}^{2} (x) =\frac{3x^{4}}{8}+\frac{x^{2}}{2}.
\label{bes_sum_examples}
\end{equation}

\section{Counting Terms Where $G(\mathbf{j}_q)=G(\mathbf{j}_q^\prime)$}
\label{appdiceprob}
To find the number of terms where $G(\mathbf{j}_q)=G(\mathbf{j}_q^\prime)$, we require that [see Eq.\ (\ref{FG2})]
\begin{equation}
( j_1 + j_2 + \ldots + j_q - j_1^\prime - j_2^\prime - \ldots - j_q^\prime) = 0, \label{jequation}
\end{equation}
where $j_q,j_q^\prime \in [0, n-1]$.  We follow a well-known number theoretical approach, which is described in detail in \cite{Apostol1976}.  The number of ways that Eq.\ (\ref{jequation}) can be satisfied is isomorphic to the problem of evaluating the $x$-independent term in 
\begin{equation}
\left( 1+x+x^2+ \ldots +x^{n-1}\right)^q \left( 1+x^{-1}+x^{-2}+ \ldots +x^{-n+1}\right)^q,
\end{equation}
which we multiply by $x^{q(n-1)}$, to give
\begin{equation}
\left( 1+x+x^2+\ldots +x^{n-1}\right)^{2q}.
\label{Eq:MultiplyIdentity}
\end{equation}
We now require the coefficient of $x^{q(n-1)}$ in Eq.\ (\ref{Eq:MultiplyIdentity}), which is given by\footnote{This is equivalent to considering $2q$ $n$-sided dice and finding the number of ways $W(2q,n)$ of totalling $q(n-1)$.} \cite{Uspensky1937}
\begin{equation}
\begin{split}
W(2q,n) = & \sum_{j=0}^q (-1)^j {2q \choose j} {n(q-j)+q-1 \choose 2q-1}  \\
=&  \sum_{j=0}^q (-1)^j {N+q-1 \choose N-q}{2q \choose j},
\end{split}
\label{W}
\end{equation}
where we have used ${x \choose y}={x \choose x-y}$ and set $N = n(q-j)$.  The only $n$-dependent part of $W(2q,n)$ is the binomial coefficient 
\begin{equation}
{N+q-1 \choose N-q} = \frac{(N+m-1)(N+m-2)\ldots  (N-m+1)}{(2q-1)!},
\end{equation}
the numerator of which is a polynomial in $N$ of degree $2q-1$.  Thus, we may write $W(2q,n)=S_{2q-1}(n)$ where $S_{2q-1}(n)$ is a polynomial in $n$ of degree $2q-1$.

\section{Derivation of moments for Gaussian distributions\label{appdist}}

\subsection{Second moment}

For a Gaussian initial momentum distribution $D(P)=\exp(-P^2/2w^2)/\sqrt{2\pi w^2}$ ($w$ is the standard deviation), the initial second-order momentum moment is $\langle \hat{P}^{2}\rangle_{0} =  w^2$.  Using Eqs.\ (\ref{p2master}) and (\ref{omega_evaluate}), we deduce that, for integer $\Omega=r$,
\begin{equation}
\expec{\hat{P}^2}_n = w^2 + \frac{\phi_d^2}{2\sqrt{2\pi w^2}}\int_{-\infty}^{\infty}dP 
\frac{\sin^{2}(n[\alpha-r]\pi/2)}{\sin^{2}([\alpha-r]\pi/2)}
e^{-P^2/2w^2}.
\label{p2gintdef}
\end{equation}
Note that we can replace the $\beta$ in $\alpha\equiv (1+2\beta)\ell$ with $P\equiv k+\beta$  without altering Eq.\ (\ref{omega_evaluate}).  We now use Eq.\ (\ref{U2}) together with
\begin{equation}
\cos(q[(1+2P)\ell-r]\pi) = (-1)^{q(\ell-r)}\cos(2q\ell\pi P),
\label{Eq:CosDoubleAngle}
\end{equation}
to determine from Eq.\ (\ref{p2gintdef}) that
\begin{equation}
\begin{split}
\expec{\hat{P}^2}_n =& w^2 + \frac{\phi_d^2}{2}n + \frac{\phi_d^2}{\sqrt{2\pi w^2}}\sum_{q=1}^{n-1}(-1)^{q(\ell-r)}(n-q)\\ &\times \int_{-\infty}^{\infty}dP\cos(2q\ell\pi P)e^{-P^2/2w^2},
\end{split}
 \label{p2gint}
\end{equation}
We now substitute  \cite{Grad2007}
\begin{equation}
\frac{1}{\sqrt{2\pi w^2}}\int_{-\infty}^{\infty}dP\cos(2q\ell\pi P)e^{-P^2/2w^2} = e^{-2q^2\ell^2\pi^2w^2},
\label{intresult1}
\end{equation}
into Eq.\ (\ref{p2gint}), which gives the final result:
\begin{equation}
\expec{\hat{P}^2}_n = w^2 + \frac{\phi_d^2}{2}n + \phi_d^2\sum_{q=1}^{n}(-1)^{q(\ell-r)}(n-q) e^{-2q^2\ell^2\pi^2w^2}.
\label{Eq:AppGauss2}
\end{equation}

\subsection{Fourth moment}

The initial fourth-order moment for a Gaussian initial distribution is $\expec{\hat{P}^4}_0=3w^4$. Using a similar approach to that for the second-order momentum moment, we use Eqs.\ (\ref{p4master}) and (\ref{omega_evaluate}) to deduce that, for $\Omega=r$,
\begin{equation}
\begin{split}
\expec{\hat{P}^{4}}_n =& 3w^{4} 
+ \frac{3\phi_d^{4}}{8\sqrt{2\pi w^2}}
\int_{-\infty}^{\infty}dP \frac{\sin^{4}(n[\alpha-\Omega]\pi/2)}{\sin^{4}([\alpha-\Omega]\pi/2)}e^{-P^2/2w^2} \\
&
+\frac{\phi_d^2}{\sqrt{2\pi w^2}}\int_{-\infty}^{\infty}dP \left(\frac{1}{2}+3P^{2}\right)
\\ & \times
\frac{\sin^{2}(n[\alpha-\Omega]\pi/2)}{\sin^{2}([\alpha-\Omega]\pi/2)}e^{-P^2/2w^2}.
\end{split}
\label{p4g_start}
\end{equation}
Using Eqs.\ (\ref{Eq:CosDoubleAngle}), (\ref{U2}) and (\ref{U4}) we deduce from Eq.\ (\ref{p4g_start}) that
\begin{equation}
\begin{split}
\expec{\hat{P}^4}=& 3w^4+ \frac{\phi_d^4}{8}(2n^2+1)n +\frac{\phi_{d}^{2}}{2}n  +3\expec{P_0^2}\phi_d^2n
\\ &+\frac{3\phi_d^4}{8\sqrt{2\pi w^2}}\sum_{q=1}^{n-1}(-1)^{q(\ell-r)} \Biggl[q^3 -2nq^2-q 
\\&
+\frac{2n}{3}(n^2+1) \Biggr]
\int_{-\infty}^{\infty}dP\cos(2q\ell\pi P)e^{-P^2/2w^2} 
\\ &- \frac{3\phi_d^4}{8\sqrt{2\pi w^2}}\sum_{q=n}^{2n-2}(-1)^{q(\ell-r)}\left[ \frac{q^3}{3}-2nq^2
+\frac{(12n^2-1)}{3}q 
\right. \\& \left. 
+\frac{2n}{3}(1-4n^2)\right]\int_{-\infty}^{\infty}dP\cos(2q\ell\pi P)e^{-P^2/2w^2} 
\\&+\frac{\phi_d^2}{\sqrt{2\pi w^2}}\sum_{q=1}^{n-1}(-1)^{q(\ell-r)}(n-q)
\\ &\times
\int_{-\infty}^{\infty}dP \cos(2q\ell\pi P)e^{-P^2/2w^2} 
\\&+\frac{6\phi_d^2}{\sqrt{2\pi w^2}}\sum_{q=1}^{n-1}(-1)^{q(\ell-r)}(n-q)
\\ &\times\int_{-\infty}^{\infty}dP P^2\cos(2q\ell\pi P)e^{-P^2/2w^2}. 
\end{split}
\label{appp4_gauss_res1}
\end{equation} 
Finally, using $\expec{\hat{P}^2}_0=w^2$, Eq.\ (\ref{intresult1}), and the integral \cite{Grad2007}
\begin{equation}
\begin{split}
\frac{1}{\sqrt{2\pi w^2}}\int_{-\infty}^{\infty}dP P^2\cos(2q\ell\pi P)e^{-P^2/2w^2} =& w^2(1-4q^2\ell^2\pi^2w^2)\\ &\times e^{-2q^2\ell^2\pi^2w^2}\label{intresult2}
\end{split}
\end{equation}
to simplify Eq.\ (\ref{appp4_gauss_res1}), we find that
\begin{equation}
\begin{split}
\expec{\hat{P}^4}=& 3w^4+ \frac{\phi_d^4}{8}(2n^2+1)n + 3w^2\phi_d^2n
\\& + \frac{3\phi_d^4}{8}\sum_{q=1}^{n-1}(-1)^{q(\ell-r)}e^{-2q^2\ell^2\pi^2w^2}
\\ &\times \left[q^3 -2nq^2-q +\frac{2n}{3}(2n^2+1)\right]
\\&-  \frac{3\phi_d^4}{8}\sum_{q=n}^{2n-2}(-1)^{q(\ell-r)}e^{-2q^2\ell^2\pi^2w^2} 
\\ &\times \left[ \frac{q^3}{3}-2nq^2+\frac{(12n^2-1)}{3}q +\frac{2n}{3}(1-4n^2)\right]
\\&+\phi_d^2\sum_{q=1}^{n-1}(-1)^{q(\ell-r)}(n-q)e^{-2q^2\ell^2\pi^2w^2}.
\\&+6\phi_d^2\sum_{q=1}^{n-1}(-1)^{q(\ell-r)}(n-q)w^2(1-4w^2q^2\ell^2\pi^2) 
\\ &\times 
e^{-2q^2\ell^2\pi^2w^2}.
\end{split}
\label{appp4_gauss_res}
\end{equation}

\section{Cosine expansions\label{appcossum}}

In order to perform many of the integrations in Appendix \ref{appdist}, we used finite cosine expansions of powers of $\sin(n\vartheta)/\sin(\vartheta)$.  To derive such an expansion for $\sin^{2}(n\vartheta)/\sin^{2}(\vartheta)$, we first consider the summation
\begin{equation}
\begin{split}
n + 2\sum_{q=1}^{n-1}(n-q)\cos(2q\vartheta)
=& n + n\sum_{q=1}^{n-1}e^{i2q\vartheta } +n\sum_{q=1}^{n-1}e^{-i2q\vartheta } \\ &- \sum_{q=1}^{n-1}qe^{i2q\vartheta } - \sum_{q=1}^{n-1}qe^{-i2q\vartheta }.
\end{split}
\label{sin2_proof}
\end{equation}
The exponential sums in Eq.\ (\ref{sin2_proof}) are geometric sums, or their derivatives, which can be evaluated to give,
\begin{equation}
\begin{split}
n + 2\sum_{q=1}^{n-1}(n-q)\cos(2q\vartheta)
=& n + n\frac{e^{i2\vartheta n}-e^{2i\vartheta }}{e^{2i\vartheta }-1} 
\\& 
-n\frac{e^{-i2\vartheta(n-1)}-1}{e^{2i\vartheta }-1}
\\& +\frac{i}{2}\frac{\partial }{\partial \vartheta }\frac{e^{i2\vartheta n}-e^{2i\vartheta }}{e^{2i\vartheta }-1} 
\\& - \frac{i}{2}\frac{\partial }{\partial \vartheta }\frac{e^{-i2\vartheta n}-e^{-2i\vartheta }}{e^{-2i\vartheta }-1}. \label{E2}
\end{split}
\end{equation}
Differentiating Eq.\ (\ref{E2}) and identifying a mutual denominator then gives the desired result:
\begin{equation}
\begin{split}
n + 2\sum_{q=1}^{n-1}(n-q)\cos(2q\vartheta)&
=\frac{e^{2i\vartheta(n+1)}-2e^{2i\vartheta}+e^{-2i\vartheta(n-1)}}{(e^{2i\vartheta}-1)^2} 
\\ & = 
\frac{\sin^2(n\vartheta)}{\sin^2(\vartheta)}.
\end{split}
\label{U2}
\end{equation}

Similarly, rewriting $\cos(2q\vartheta)$ and converting between terms involving powers of $q$ to derivatives of $\cos(2q\vartheta)$, we find that
\begin{equation}
\begin{split}
\frac{\sin^4(n\vartheta)}{\sin^4(\vartheta)} =& \frac{n}{3}(2n^2+1) \\
&+ \sum_{q=1}^{n-1}\left[q^3-2nq^2-q+\frac{2n}{3}(2n^2+1) \right]\cos(2q\vartheta) \\
&-\sum_{q=n}^{2n-2}\left[\frac{q^3}{3}-2nq^2 + (4n^2-\frac{1}{3})q + \frac{2n}{3}(1-4n^2) \right] \\ &\times \cos(2q\vartheta). \label{U4}
\end{split}
\end{equation}

We note that $U_{n-1}(\cos(\vartheta))=\sin(n\vartheta)/\sin(\vartheta)$ is a Chebyshev polynomial of the second kind, and in this appendix we have considered finite expansions of these in terms of Chebyshev polynomials of the first kind, i.e., $T_n(\cos(\vartheta ))=\cos(n\vartheta )$.  These are equivalent to discrete Fourier transforms.
\end{appendix}

\newpage
\bibliography{Cumulant_latest}

\end{document}